\documentclass[12pt,letter]{article}
\pdfoutput=1
\usepackage{graphicx, epsfig, color,cite}
\usepackage{amsmath}
\usepackage{amssymb}
\usepackage{float}
\usepackage{hyperref}
\usepackage{subcaption}

\def\to{\rightarrow}

\def\bi{\begin{itemize}}
\def\ei{\end{itemize}}
\def\te{\tilde e}
\def\tchi{\tilde\chi}

\def\tu{\tilde u}

\def\tb{\tilde b}

\def\tst{\tilde t}
\def\ttau{\tilde \tau}

\def\tg{\tilde g}
\def\tnu{\tilde\nu}

\def\alt{\lesssim}
\def\agt{\gtrsim}
\def\be{\begin{equation}}  
\def\ee{\end{equation}}  
\def\bea{\begin{eqnarray}}  
\def\eea{\end{eqnarray}}

\begin{document}
\begin{titlepage}
\begin{flushright}
OU-HEP-241118
\end{flushright}

\vspace{0.5cm}
\begin{center}
  {\Large \bf Living dangerously with decoupled first/second\\
    generation scalars: SUSY prospects at the LHC}\\
\vspace{1.2cm} \renewcommand{\thefootnote}{\fnsymbol{footnote}}
{\large Howard Baer$^{1}$\footnote[1]{Email: baer@ou.edu },
Vernon Barger$^2$\footnote[2]{Email: barger@pheno.wisc.edu} and
Kairui Zhang$^1$\footnote[5]{Email: kzhang25@ou.edu}
}\\ 
\vspace{1.2cm} \renewcommand{\thefootnote}{\arabic{footnote}}
{\it 
$^1$Homer L. Dodge Department of Physics and Astronomy,
University of Oklahoma, Norman, OK 73019, USA \\[3pt]
}
{\it 
$^2$Department of Physics,
University of Wisconsin, Madison, WI 53706 USA \\[3pt]
}

\end{center}

\vspace{0.5cm}
\begin{abstract}
\noindent
The string landscape statistical draw to large scalar soft masses leads
to a mixed quasi-degeneracy/decoupling solution to the SUSY flavor and CP
problems where first/second generation matter scalars lie in the 20-40 TeV
range.
With increasing first/second generation scalars, SUSY models
actually become more natural due to two-loop RG effects which suppress
the corresponding third generation soft masses.
This can also lead to substantial parameter space regions which are
forbidden by the presence of charge and/or color breaking (CCB) minima of the
scalar potential.
We outline the allowed SUSY parameter space for the
gravity-mediated three extra-parameter-non-universal Higgs model NUHM3.
The natural regions with $m_h\sim 125$ GeV, $\Delta_{EW}\alt 30$ and
decoupled first/second generation scalar are characterized by
rather heavy gluinos and EW gauginos, but with rather small $\mu$ and
top-squarks not far beyond LHC Run 2 limits.
This scenario also explains why SUSY has so far eluded discovery at LHC
in that the parameter space with small scalar and gaugino masses is 
all excluded by the presence of CCB minima.

\end{abstract}
\end{titlepage}

\section{Introduction}
\label{sec:intro}

At the end of the twentieth century, it was commonly believed that weak scale
supersymmetry (SUSY) would be revealed at sufficiently energetic colliding
beam experiments with sparticle masses in the few hundred GeV
range\cite{Murayama:2000dw}.
The dominant SUSY breaking paradigm at the time was that of gravity mediation,
where hidden sector SUSY breaking would be communicated to the visible sector  
via gravitational couplings already present in the supergravity
Lagrangian\cite{Cremmer:1982en,Arnowitt:2012gc}. 
While SUSY provided a 't Hooft technical naturalness solution to the big
hierarchy problem\cite{Witten:1981nf,Kaul:1981wp},
motivations from practical naturalness\cite{Baer:2015rja}--
that SUSY contributions to the weak scale should be comparable to or less
than the weak scale-- seemingly pointed to a few hundred GeV as the SUSY
breaking mass scale\cite{Barbieri:1987fn}.
In spite of this, there were already conflicting indications that the SUSY
breaking scale could well be higher, in the TeV or beyond range.
At issue was that in generic SUGRA theories, large flavor-violating and
CP-violating effects were expected as there was no known symmetry to
suppress these.
In fact, the SUSY flavor and CP problems soon motivated alternative
paradigms-- gauge mediation\cite{Dine:1995ag}, anomaly mediation\cite{Randall:1998uk} and
gaugino mediation\cite{Schmaltz:2000gy}-- which led to universal scalar soft
terms and suppression of flavor violating effects via a super-GIM mechanism. 

However, already in 1993 Dine et al.\cite{Dine:1993np} proposed an alternative
for SUGRA-based theories that the first two generations of scalar masses could
lie in the tens of TeV range, yielding a decoupling solution to the flavor
and CP problems, whilst third generation scalar masses could lie in the
TeV-range, which was more consistent with naturalness.
This proposal was further amplified by Cohen et al.\cite{Cohen:1996vb} (CKN)
with their 'more minimal' or effective SUSY scheme which delineated the
attractive features of such a scenario\cite{Baer:2010ny}.
This also motivated the radiatively-generated scalar mass hierarchy models of
Bagger et al.\cite{Bagger:1999sy} where all scalars had multi-TeV values
at the high scale, but where large (unified) Yukawa coupling effects drive
the third generation masses to TeV-scale values while leaving first/second
generation scalars in the tens-of-TeV range\cite{Baer:2000jj}. 

In the 21st century, the discovery of $m_h\sim 125$ GeV seemed to require for
SUSY theories either large trilinear soft ($A$) terms or else top squark
soft terms in the $\sim 10-100$ TeV range when $A$-terms were small
(as was expected for gauge, anomaly or gaugino mediation)\cite{Arbey:2011ab}. 
For this latter case, the large third generation scalar masses violated even
the most conservative naturalness conditions, thus selecting out
gravity-mediation models as being able to generate $m_h\sim 125$ GeV whilst preserving naturalness and maintaining consistency with LHC sparticle mass
search limits\cite{Baer:2014ica}.
But if gravity-mediation was then preferred, what of the SUSY flavor and
CP problems?\cite{Gabbiani:1996hi}
One possibility was to abandon naturalness as in
split\cite{Arkani-Hamed:2004ymt,Arkani-Hamed:2004zhs} or
minisplit\cite{Arvanitaki:2012ps}  SUSY models which could then allow
for the required 10-100 TeV third generation scalars\cite{Arkani-Hamed:1997opn}.
But in these models, rather light gauginos and higgsinos were also
expected and so these models could not escape pressure from negative
LHC searches.

\section{Inverted scalar mass hierarchy from the string landscape}
\label{sec:land}

Another route to sparticle spectra with an inverted scalar mass hierarchy
($m_0(1,2)\gg m_0(3)$) arose with the advent of the string landscape.
Under flux compactifications of type II-B string theory\cite{Bousso:2000xa,Douglas:2006es}, the huge number of compactification possibilities,
of order $10^{500-272,000}$\cite{Ashok:2003gk}, each leads to different $4-d$
laws of physics.
This scenario provides a basis for Weinberg's anthropic solution to the
cosmological constant problem\cite{Weinberg:1987dv}:
in the greater multiverse with a huge number of vacua with the SM as low
energy EFT, but with different cosmological constants, then it may be no
surprise that we find ourselves in a pocket universe with
$\rho_{vac}\sim 10^{-122}m_P^4$ since if it were much bigger, the universe
would expand so quickly that structure, and hence observers, would not arise. 

This logic can also be applied to other dimensionful quantities, such as the
overall SUSY breaking scale\cite{Denef:2004ze,Douglas:2004qg},
which in turn determines the magnitude of the weak scale.
The different functional dependence of the various soft SUSY breaking terms
on hidden sector/moduli fields leads them to effectively scan independently
in the multiverse\cite{Baer:2020vad}. 
Since nothing in string theory favors one over another value of SUSY breaking
scales, then the $F$-breaking fields are expected to be uniformly distributed
as complex numbers whilst the $D$-breaking fields scan as real numbers.
The overall SUSY breaking scale is then statistically expected to scan as a
power-law draw to large terms as\cite{Douglas:2004qg,Susskind:2004uv,Arkani-Hamed:2005zuc}
\be
f_{SUSY}\sim m_{soft}^{2n_F+n_D-1}
\ee
where $m_{soft}\sim m_{hidden}^2/m_P$, and $n_F$ is the number of $F$-breaking
fields and $n_D$ is the number of $D$-breaking fields contributing to the
overall SUSY breaking scale.
Thus, even for the textbook case of SUSY breaking via a single $F$-term field,
the multiverse would favor a statistical linear draw to large soft terms,
and if more SUSY breaking fields also contribute, then the draw to large
soft terms would be even stronger.

Naively, this picture would favor SUSY breaking at the highest scales
possible, but now one must fold in anthropics.
Restricting ourselves to string vacua with the MSSM as low scale EFT
(as expected in Calabi-Yao compactifications\cite{Candelas:1985en}),
then the weak scale is determined by minimizing the scalar potential.
The $Z$ mass in each pocket universe is given by
\be
(m_Z^{PU})^2/2=\frac{m_{H_d}^2+\Sigma_d^d-(m_{H_u}^2+\Sigma_u^u)\tan^2\beta}{\tan^2\beta -1}-\mu^2\simeq-m_{H_u}^2-\mu^2-\Sigma_u^u(\tst_{1,2}) .
\label{eq:mzsPU}
\ee
Here, $m_{H_u}^2$ and $m_{H_d}^2$ are the Higgs field soft terms, $\mu$ is the
SUSY conserving $\mu$ parameter\footnote{Twenty solutions to the SUSY
  $\mu$ problem are reviewed in Ref. \cite{Bae:2019dgg}.},
$\tan\beta =v_u/v_d$ is the ratio of Higgs field vevs and the 
$\Sigma_u^u$ and $\Sigma_d^d$ contain over 40 loop corrections, which can be
found in the Appendix to Ref. \cite{Baer:2012cf}, using the Coleman-Weinberg
formalism\cite{Coleman:1973jx}.
The anthropic condition is that $m_Z^{PU}$ must lie not-to-far from its
measured value in our universe, lest the up-down quark mass difference lead
to pocket universes with either all protons, or all neutrons,
and no complex nuclei, which are needed for observers.
This window of anthropically-allowed weak
scale values is called the ABDS window\cite{Agrawal:1997gf}:
\be
0.5 m_Z^{OU}\alt m_Z^{PU}\alt (2-5)m_Z^{OU}
\ee
where $m_Z^{OU}=91.2$ GeV.
Thus, if a contributing soft term is too large, it will 
typically lead to contributions to Eq. \ref{eq:mzsPU} beyond the ABDS window,
unless some implausible accidental finetuning occurs.

In most landscape scans with small $\mu\sim 100-350$ GeV,
the $\Sigma_u^u(\tst_{1,2})$ are the largest weak scale contributions
since their Yukawa couplings are large:
\be
\Sigma_u^u(\tst_{1,2})=\frac{3}{16\pi^2}F(m_{\tst_{1,2}})\times\left[ f_t^2-g_Z^2\mp\frac{f_t^2A_t^2-8g_Z^2(\frac{1}{4}-\frac{2}{3}x_W)\Delta_t}{m_{\tst_2}^2-m_{\tst_1}^2}\right]
\label{eq:Sigmauu}
\ee
where $\Delta_t=(m_{\tst_L}^2-m_{\tst_R}^2)/2+m_Z^2\cos 2\beta(\frac{1}{4}-\frac{2}{3}x_W)$, 
$g_Z^2=(g^2+g^{\prime 2})/8$, $x_W\equiv\sin^2\theta_W$ and $F(m^2)=m^2(\log (m^2/Q^2)-1)$ with
$Q^2\simeq m_{\tst_1}m_{\tst_2}$.
Hence, top-squarks cannot be too heavy.
Since $\Sigma_u^u(\tst_{1,2})\sim m_{\tst_1}^2/16\pi^2$, the top squark masses
lie typically in the few TeV range.
(In addition, for large $A_t$ terms, there can be large cancellations
in the numerator of $\Sigma_u^u$.\cite{Baer:2012up})
However, for first/second generation squarks and sleptons, since their
Yukawa couplings are much smaller, then they would be selected to be
far heavier, in the 10-40 TeV range\cite{Baer:2017uvn}. 
Actually, the upper bounds on first/second generation sfermions do not come
from the $\Sigma_u^u$ terms but rather from 2-loop RG running\cite{Martin:1993zk} of third
generation sfermions\cite{Arkani-Hamed:1997opn,Baer:2000xa}:
\be
\frac{dm_i^2}{dt}=\frac{1}{16\pi^2}\beta_{m_i^2}^{(1)}+\frac{1}{(16\pi^2)^2}\beta_{m_i^2}^{(2)} ,
\label{eq:dmi}
\ee
where $t=\log (Q)$ and $i=Q_j$, $U_j$, $D_j$, $L_J$ and $E_j$ with $j=1-3$ a generation index.
The $\beta_{m_i^2}^{(1)}$ are the usual 1-loop beta functions while the 2-loop 
beta functions contain the terms
\be
\beta_{m_i^2}^{(2)}\ni a_i g_3^2\sigma_3+b_i g_2^2\sigma_2+c_ig_1^2\sigma_1
\ee
and where
\bea
\sigma_1 &=& \frac{1}{5}g_1^2\left\{3(m_{H_u}^2+m_{H_d}^2)+\text{Tr}[{\bf m}_{Q}^2+3{\bf m}_L^2+8{\bf m}_U^2+2{\bf m}_D^2+2{\bf m}_E^2]\right\} \label{eq:sigmas_be} \\
\sigma_2 &=& g_2^2\left\{ m_{H_u}^2+m_{H_d}^2 +\text{Tr}[3{\bf m}_Q^2+{\bf m}_L^2 ]\right\} \\
\sigma_3 &=& g_3^2 \text{Tr}[2{\bf m}_Q^2+{\bf m}_U^2+{\bf m}_D^2 ] .
\label{eq:sigmas_ee}
\eea
The coefficients $a_i$, $b_i$ and $c_i$ are related to the quantum numbers of
the scalar fields but are all positive constants.
Normally, the 2-loop contributions to Eq. \ref{eq:dmi} are small
relative to the 1-loop contributions.
But the $\sigma_i$ terms contain all scalar masses, so when some of these
become huge, then the 2-loop terms can be comparable or even exceed the
1-loop beta functions.
In our case with $m_0(1,2)\gg m_0(3)$, then the large
first/second generation scalar masses leads to {\it suppression} of 3rd
generation scalar masses in their running from $Q=m_{GUT}$ to $Q=m_{weak}$,
and thus typically a diminution of the $\Sigma_u^u(\tst_{1,2})$ terms:
{\it i.e.} the theory becomes more {\it natural}\cite{Baer:2019zfl}!
Since the terms in Eq's \ref{eq:sigmas_be}-\ref{eq:sigmas_ee} depend on gauge and not Yukawa
couplings, the upper bounds on first/second generation sfermions masses are
the same, leading to a  landscape selection of quasi-degeneracy,
{\it i.e.} degeneracy at the $\sim 1-20\%$ level\cite{Baer:2019zfl}.

\begin{figure}[htb!]
\centering
    {\includegraphics[height=0.4\textheight]{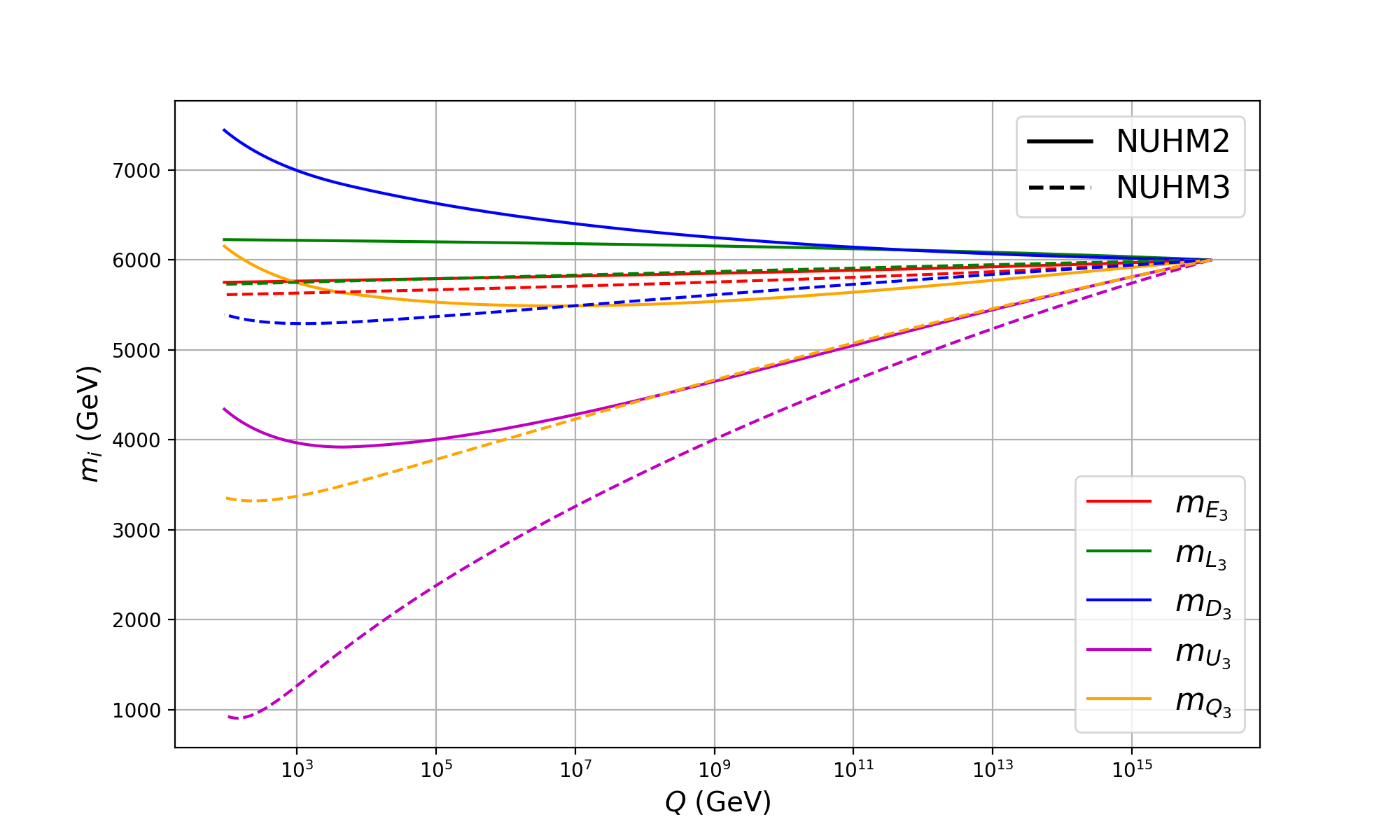}}
    \caption{Running third generation soft scalar masses vs. scale $Q$ in the NUHM2 model
listed in Table \ref{tab:bm} with $m_0(1,2)=6$ TeV (solid) and NUHM3 (dashed) with 
$m_0(1,2)=30$ TeV. We also take
$A_0=-m_0(3)$ TeV, $m_0(3)=6$ TeV, $m_{1/2}=2.2$ TeV,
      $m_A=2$ TeV, $\mu =200$ GeV and $\tan\beta =10$.
                \label{fig:msoft}}
\end{figure}

The third generation scalar soft term running is shown in Fig. \ref{fig:msoft}
for $m_0(3)=6$ TeV, $m_{1/2}=2.2$ TeV, $A_0=-m_0(3)$, $\tan\beta =10$ with
$\mu =200$ GeV and $m_A=2000$ GeV.
For the solid curves in NUHM2, we take scalar mass degeneracy with
$m_0(1,2)=m_0(3)$ while for the dashed curves in NUHM3 we take
$m_0(1,2)=30$ TeV.
From the plot, the downward RG push on all five third generation scalar masses
is evident and  $m_{U_3}^2$ is particularly noticeable in that its
weak scale value shifts from $\sim 4$ TeV for universality to
$m_{U_3}\sim 1$ TeV for $m_0(1,2)=30$ TeV.
The downward push on third generation masses can reduce the dominant
$\Sigma_u^u (\tst_{1,2})$ terms in Eq. \ref{eq:mzsPU}, making the spectra
more natural.
If the value of $m_0(1,2)$ is increased even more beyond $\sim 30$ TeV, then
in fact the $m_{U_3}^2$ soft term can be driven to tachyonic values leading to
CCB minima in the scalar potential.
Such CCB minima must also be vetoed in the multiverse as leading to
unlivable universes.
%

\section{Landscape scan of NUHM3 model}
\label{sec:scan}

Here, we scan the landscape of NUHM3 vacua over the range\footnote{We use Isajet 7.91\cite{Paige:2003mg} for
sparticle and Higgs mass and $\Delta_{EW}$ generation.}
We scan over the following parameters but with a {\it linear} $n=1$ selection
favoring large soft terms:
\bi
\item $m_0(1, 2): 20-50$ TeV,
\item $m_0(3): 0-15$ TeV,
\item $m_{1/2}: 0-3.5$ TeV,
\item $A_0: 0- (-25)$ TeV (negative only),
\item $m_A: 0-10$ TeV,
\item $\tan\beta: 3-60$ (flat) and
  \item $\mu =200$ GeV.
\ei
for fixed $\mu =200$ GeV (since $\mu$ is not a soft term but arises
from whatever solution to the SUSY $\mu$ problem is assumed).
For each parameter set, the expected value of the weak scale in each pocket
universe is computed from $m_{weak}^{PU}=m_Z\sqrt{\Delta_{EW}/2}$ and
solutions with $m_{weak}^{PU}> 4m_{weak}^{OU}$ are rejected as lying beyond the
ABDS window. (An important technical point is that the scan range upper limits
must be selected to be beyond the upper limits imposed by the anthropic
ABDS window.) To gain the probability distributions for Higgs and sparticle
masses in {\it our} universe, then as a final step $m_Z^{PU}$ must be adjusted
(but not finetuned!) to our value $m_Z^{OU}=91.2$ GeV.

\subsection{Distribution of $m_0(1)$, $m_0(2)$ and $m_{\tst_1}$ from the landscape}
\label{ssec:scan}

\begin{figure}[htb!]
\centering
    {\includegraphics[height=0.4\textheight]{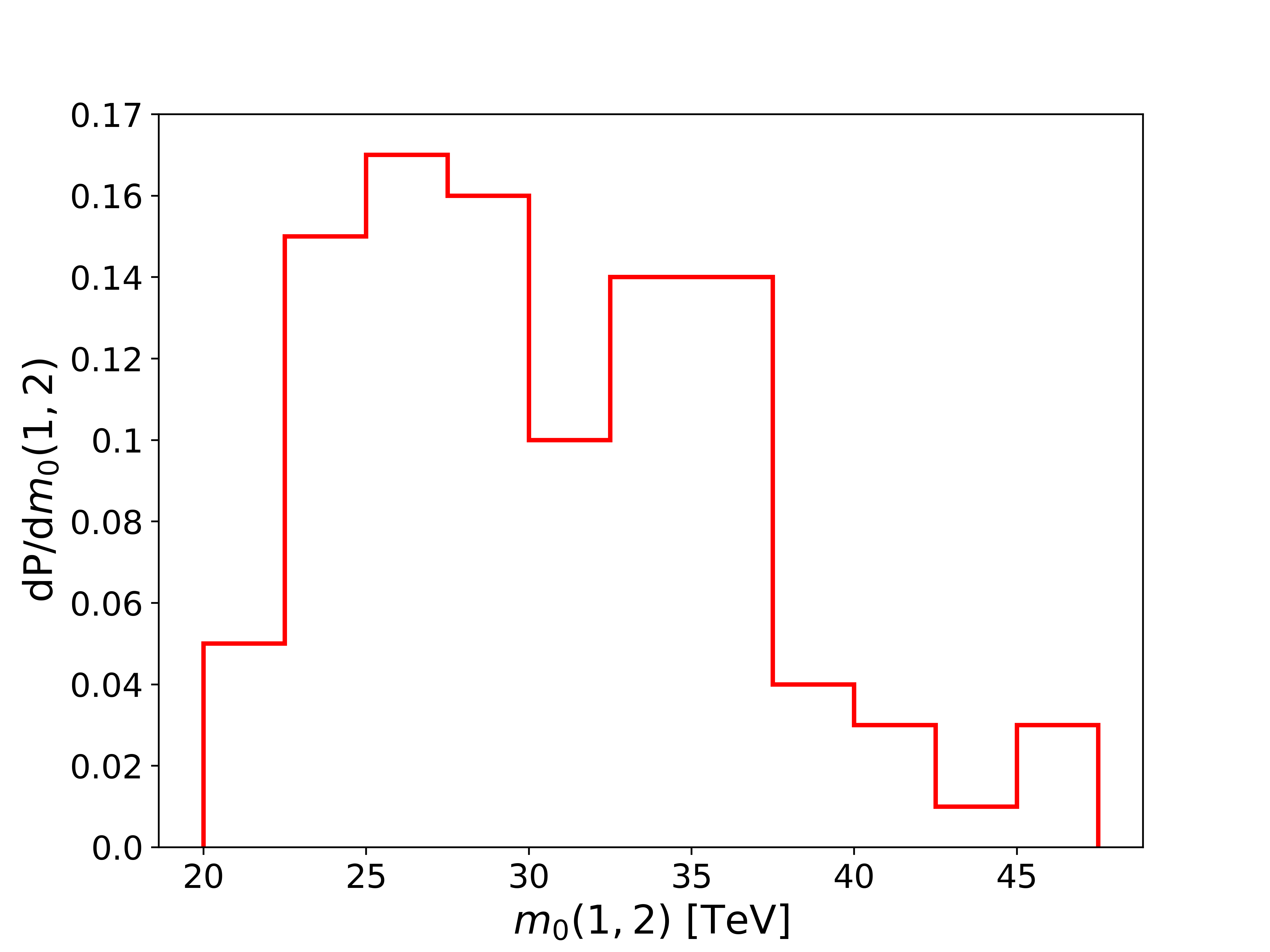}}\\
        {\includegraphics[height=0.4\textheight]{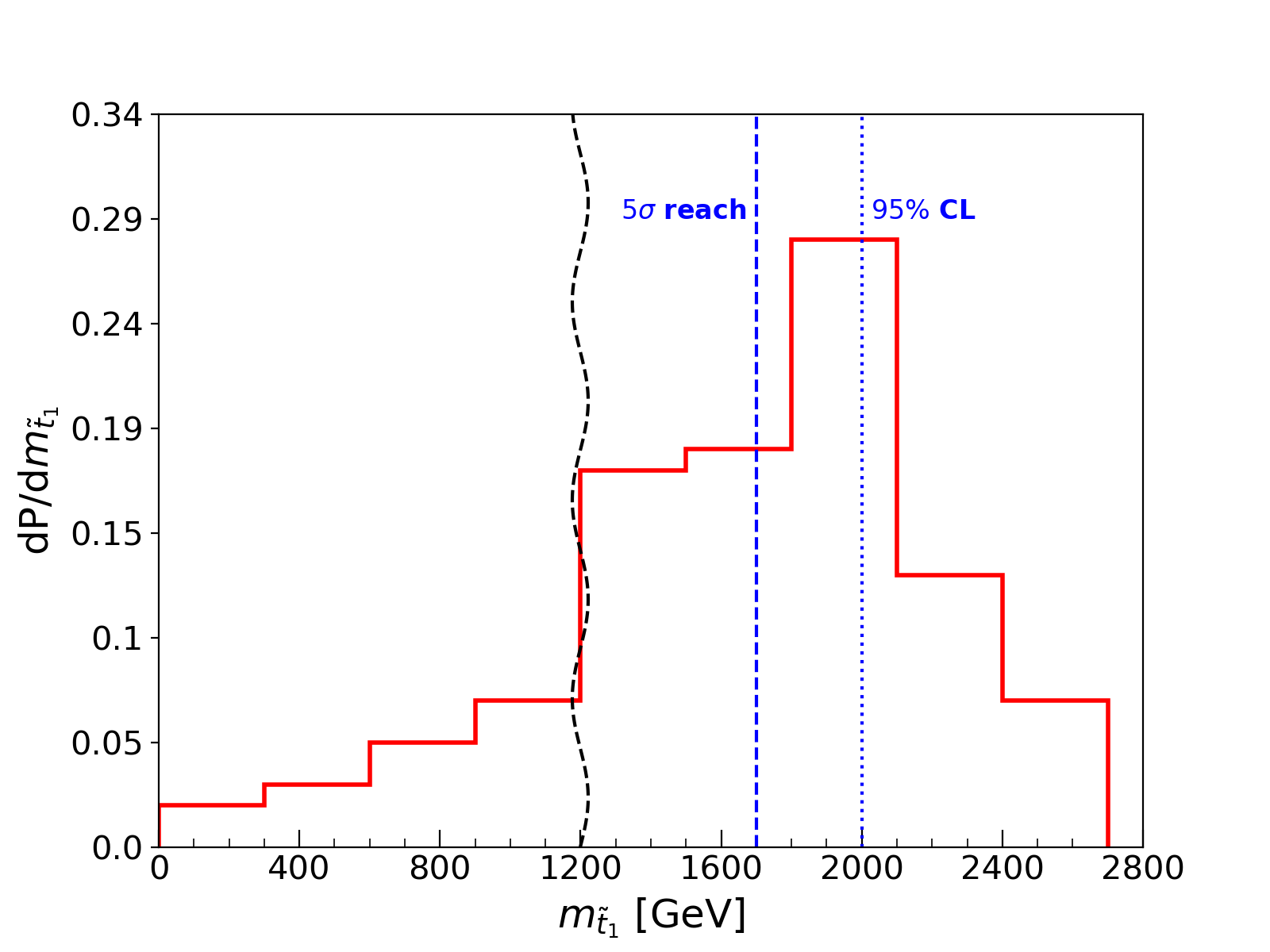}}
    \caption{Distribution in {\it a}) $dP/dm_0(1,2)$ vs $m_0(1,2)$
and {\it b}) $dP/dm_{\tst_1}$ vs. $m_{\tst_1}$
for a landscape scan in the NUHM3 model.
The wavy line in frame {\it b}) corresponds to current LHC Run 2 limits
on $m_{\tst_1}$ from simplified model searches whilst the other vertical lines correspond to the projected stop mass reach of HL-LHC within natural SUSY models.
      \label{fig:m012}}
\end{figure}

The resulting probability distribution for {\it the average} first/second
generation scalar masses is shown in Fig. \ref{fig:m012}{\it a}).
From the plot, we see that $m_0(1,2)$ actually has a broad peak from
$\sim 25-35$ TeV which then tails off towards zero around 45 TeV.
Since we expect independent (non-universal ) generations from
gravity-mediation, then $m_0(1,2) $ is interpreted as an 
{\it average} of $m_0(1)$ and $m_0(2)$  but where $m_0(1)\sim m_0(2)$:
partially decoupled with quasi-degeneracy which as shown in
Ref. \cite{Baer:2019zfl} can resolve the SUSY flavor and CP problems.

In Fig. \ref{fig:m012}{\it b}), we show the corresponding landscape
scan in $m_{\tst_1}$. In this case, the range of $m_{\tst_1}$ starts at
below the 1 TeV level but then builds to a peak probability distribution
arond 2 TeV before tapering off at around 3 TeV. We also show the current ATLAS/CMS rough bound on $m_{\tst_1}$ from simplified model searches for top-squark
pair production (wavy line) and the $5\sigma$ and 95\% CL reach of HL-LHC with
$\sqrt{s}=14$ TeV and 3000 fb$^{-1}$\cite{Baer:2023uwo}.

\subsection{Improved naturalness from decoupled first/second generation scalars}

\begin{figure}[htb!]
\centering
    {\includegraphics[height=0.4\textheight]{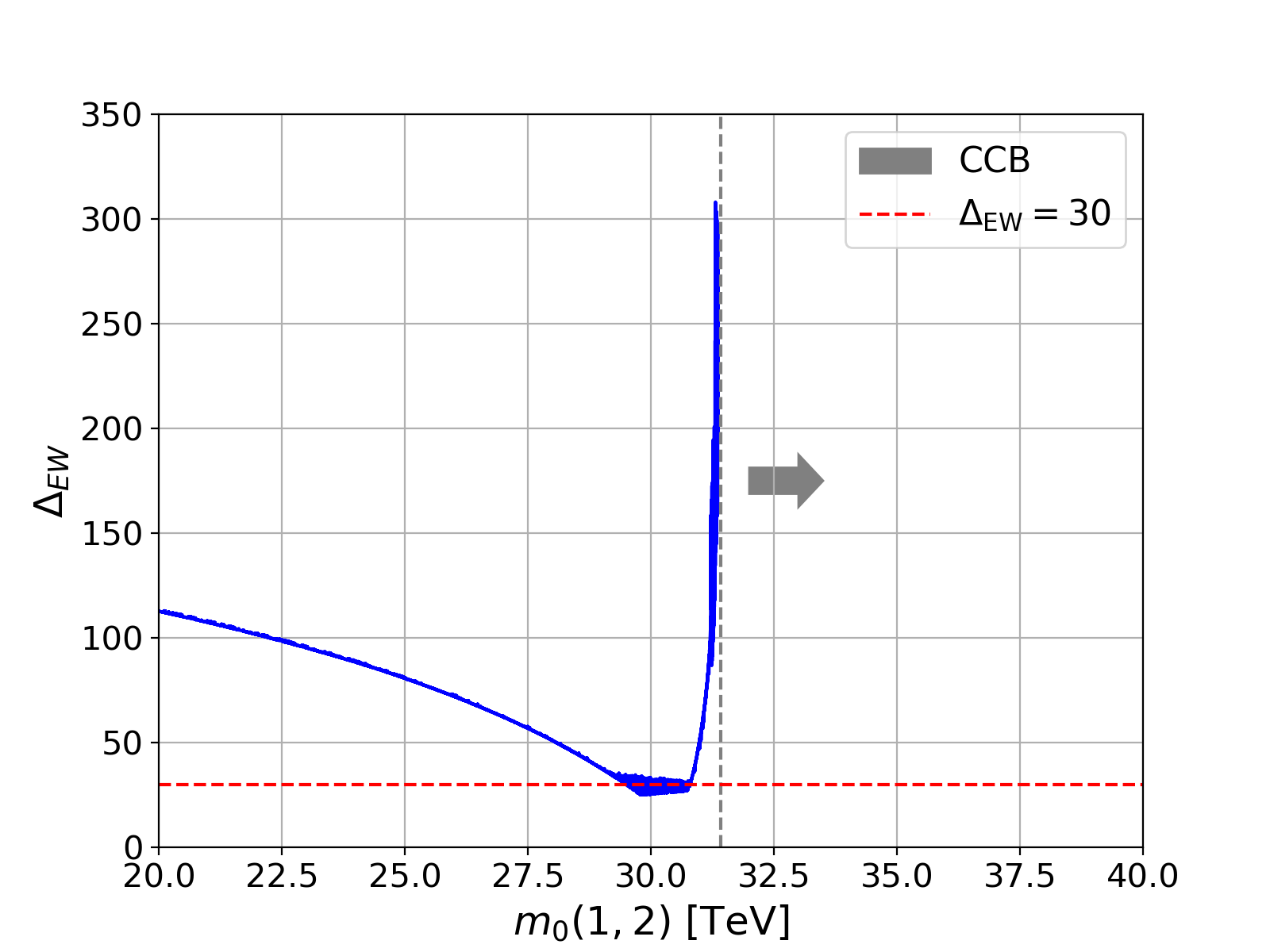}}
    \caption{Naturalness measure $\Delta_{EW}$ vs. $m_0(1,2)$
      for $A_0=-m_0(3)$ TeV, $m_0(3)=6$ TeV, $m_{1/2}=2.2$ TeV,
      $m_A=2$ TeV, $\mu =200$ GeV and $\tan\beta =10$.
                \label{fig:dewvsm0}}
\end{figure}

The behavior of improved naturalness arising from decoupled first/second
generation scalars is shown in Fig. \ref{fig:dewvsm0}, where we plot
$\Delta_{EW}$ vs. increasing $m_0(1,2)$ for the same NUHM3 parameters as used
in Fig. \ref{fig:msoft}.
Starting with $m_0(1,2)\sim 20 $ TeV, we see that $\Delta_{EW}\sim 100$,
and is dominated  by the $\Sigma_u^u (\tst_{1,2})$ terms since we have
chosen $\mu=200$ GeV.
But as $m_0(1,2)$ increases to even larger values, the top squark soft
masses become RG suppressed by the 2-loop terms, and so $\Delta_{EW}$
steadily decreases to $\Delta_{EW}\alt 30$ for $m_0(1,2)\sim 30$ TeV.
For even higher values of $m_0(1,2)$, then the lighter top squark $\tst_1$
drops into  the LHC-excluded region with $m_{\tst_1}\ll 1.1$ TeV, and
furthermore, previous large cancellations in the numerator of
Eq. \ref{eq:Sigmauu} no longer occur, so $\Delta_{EW}$ shoots 
upwards to large values shortly before the spectra become tachyonic
and CCB minima occur.
Thus, even higher values of $m_0(1,2)$ are ruled out.
This behavior of soft SUSY breaking parameters is an example of what
Arkani-Hamed {\it et al.} (ADK)\cite{Arkani-Hamed:2005zuc} label as
{\it living dangerously} on the landscape in
that some parameter or quantity can be drawn to such large values that the
pocket universe is on the edge of catastrophic breakdown: in this case
soft terms are pulled just beyond the edge of their range that
would lead to the onset of catastrophic CCB minima. 

\begin{figure}[htb!]
    \centering
    \begin{subfigure}{\textwidth}
        \centering
        \includegraphics[height=0.3\textheight]{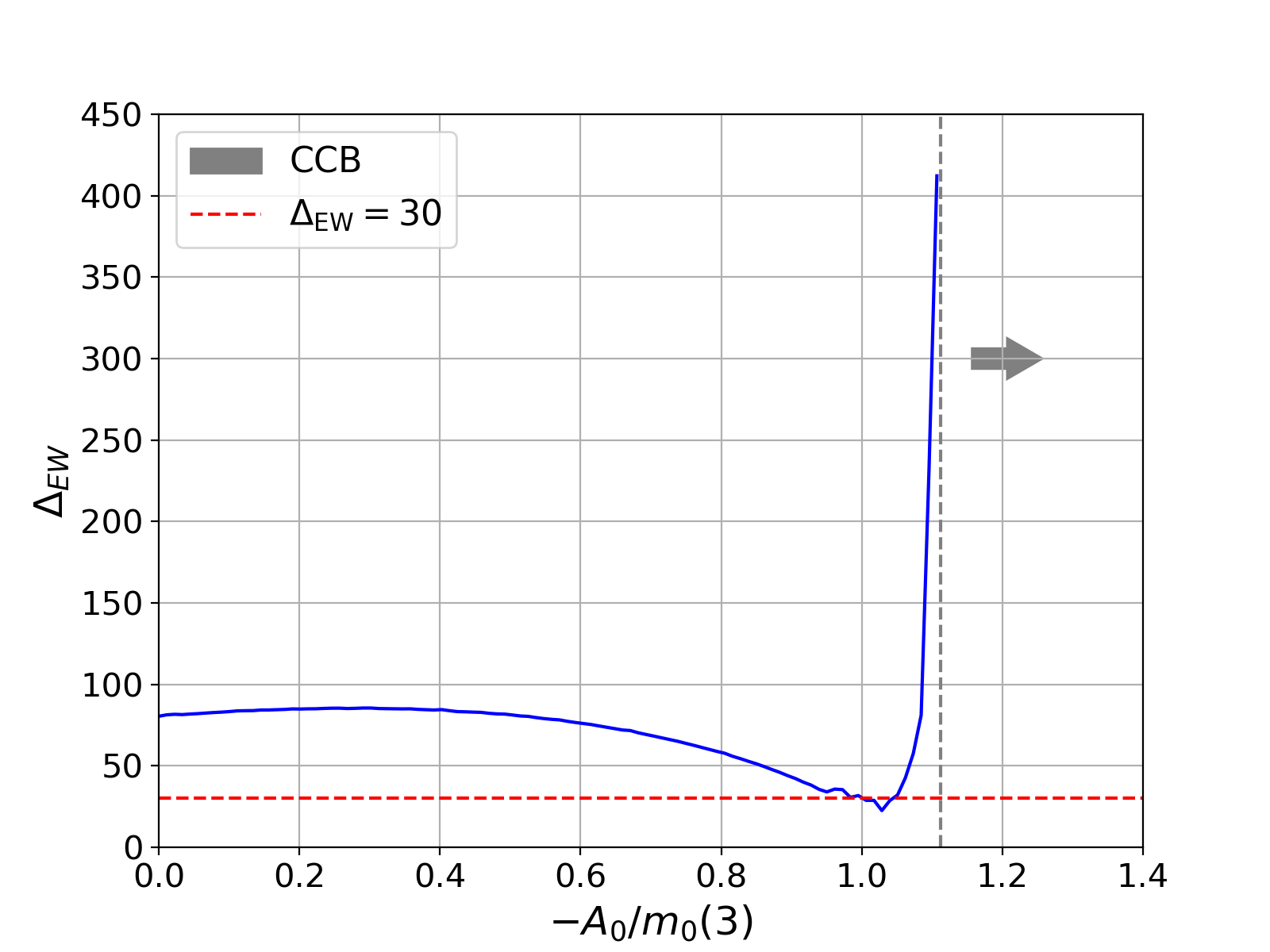}
        \caption{}
    \end{subfigure}
    \begin{subfigure}{\textwidth}
        \centering
        \includegraphics[height=0.3\textheight]{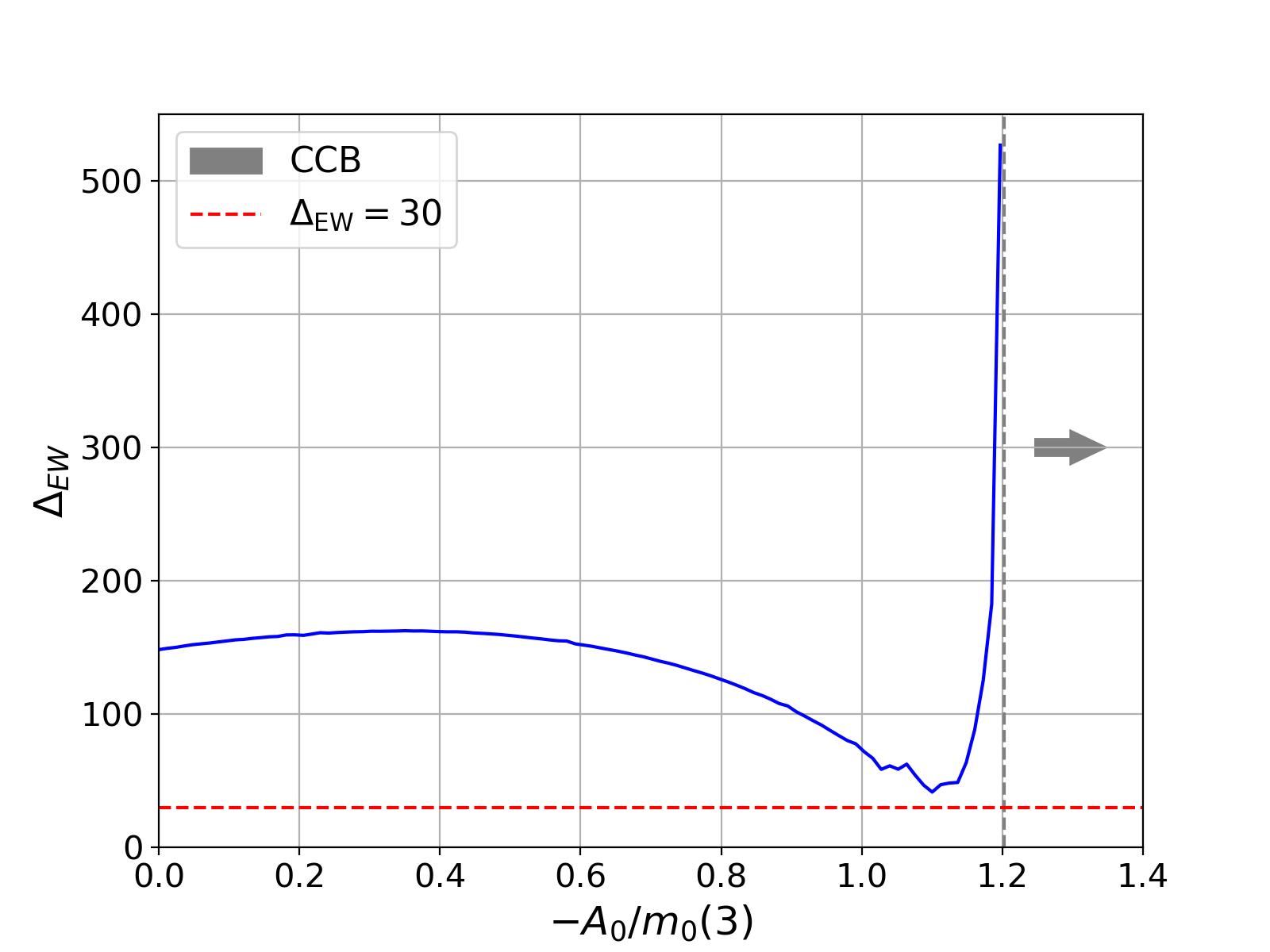}
        \caption{}
    \end{subfigure}
    \caption{Naturalness measure $\Delta_{EW}$ vs. $-A_0/m_0(3)$
      for {\it a}) $m_0(1,2)=30$ TeV, $m_0(3)=5.85$ TeV, $m_{1/2}=2.25$ TeV
      and {\it b}) $m_0(1,2)=40$ TeV, $m_0(3)=8$ TeV, $m_{1/2}=3$ TeV.
      For both frames, we take $m_A=2$ TeV, $\mu =200$ GeV and $\tan\beta =10$.
                \label{fig:dewvsA0}}
\end{figure}

Similar behavior can be found in the landscape draw of the trilinear soft
term $A_t$ to large (negative) values. In decades past, it was frequent to
see the SUSY parameter space plotted in the $m_0$ vs. $m_{1/2}$ parameter
space of models like the CMSSM, 
which was expected to be a manifestation of gravity mediation.
These plots would often be shown for $A_0=0$, which allowed one to display
(among other things) the focus point region\cite{Feng:1999mn}
of CMSSM parameter space, which occurs when $A_0\sim 0$.
However, unless one has in hand a specific mechanism to suppress $A_0$,
this goes against supergravity expectations, where 
{\it all} soft terms are expected of order the gravitino mass $m_{3/2}$. 
Furthermore, using naturalness measures such as
$\Delta_{HS}\equiv \delta m_{H_u}^2/m_{H_u}^2(\Lambda )$\cite{Kitano:2006gv},
it was expected that large $A$-terms would lead to {\it increased}  finetuning.
However, in Fig. \ref{fig:dewvsA0} we show again the EW finetuning parameter
as derived for the set of fixed NUHM3 model parameters versus $-A_0/m_0(3)$
with {\it a}) $m_0(1,2)=30$ TeV and {\it b}) $m_0(1,2)=40$ TeV.
Variation of the strength of the $A_0$ parameter can lead to 
changes in $\Sigma_u^u(\tst_{1,2})$ in three different ways:
1. larger $A_0$ values influence the RG running of third generation soft
masses, and can {\it decrease} their weak scale values if taken large enough,
2. large $A_0$ and consequently large $A_t(weak)$ influences top-squark mixing
and can lead to large splitting of the stop mass eigenstates that enter
Eq. \ref{eq:Sigmauu} and
3. large $A_t(weak)$ can lead to large cancellations in the numerator
of Eq. \ref{eq:Sigmauu}, thus also leading to greater naturalness due to
smaller $\Sigma_u^u$  values.
This behavior is realized in Fig. \ref{fig:dewvsA0}{\it a}) where
$\Delta_{EW}$ starts  off at $\sim 80$ for $A_0\sim 0$.
As $A_0/m_0(3)$ increases, there is a slight increase in $\Delta_{EW}$
followed by a slide to low values with $\Delta_{EW}$ reaching below
$\sim 30$ for $-A_0/m_0(3)\sim 1$.
For even higher $A_0$ values, the cancellations in $\Sigma_u^u$ are upset, 
and $\Delta_{EW}$ shoots upward shortly before driving stop soft terms to
tachyonic values and thus CCB minima in the scalar potential.
Similar behavior is shown in frame {\it b}) with $m_0(1,2)\sim 40 $ TeV.
In this case, the onset of CCB minima occurs for $-A_0/m_0(3)\sim 1.2$.

In Table \ref{tab:bm}, we show spectra for the two SUSY benchmark cases
illustrated in Fig. \ref{fig:msoft} with $m_0(3)=6$ TeV, $m_{1/2}=2.2$ TeV,
$\tan\beta =10$, $A_0=-6$ TeV and with $\mu =200$ GeV and $m_A=2$ TeV.
For the NUHM2 case, we take $m_0(1,2)=6$ TeV (universality) while for
NUHM3 we take $m_0(1,2)=30$ TeV.
For the NUHM2 BM point, we see that $m_h=123.3$ GeV with $m_{\tst_1}\simeq 4$
TeV. The value of $\Delta_{EW}=158$, so the model is finetuned in the electroweak.
Proceeding to the NUHM3 BM point with $m_0(1,2)=30$ TeV, and all else the same
as for NUHM2, then we see that $m_{\tst_1}$ has dropped to $m_{\tst_1}\sim 1$ TeV,
while $m_h$ has increased to 124.5 GeV and $\Delta_{EW} =25$, so the model
apears now to be EW natural!
For the NUHM2 case, only light higgsino pair production will be available to
HL-LHC searches.
For the more natural NUHM3 BM point, which allows for the mixed
decoupling/quasi-degeneracy solution to the SUSY flavor and CP problems,
then also top-squark pair production should be available, and is in fact
just below the current ATLAS/CMS mass limits for top-squark pair production
within simplied models\cite{ATLAS:2020dsf,ATLAS:2020xzu,CMS:2021beq}.
%
\begin{table}\centering
\begin{tabular}{lcc}
\hline
parameter & NUHM2 & NUHM3 \\
\hline
$m_0(1,2)$      & 6000 & 30000 \\
$m_0(3)$      & 6000 & 6000  \\
$m_{1/2}$      & 2200 & 2200 \\
$\tan\beta$    & 10 & 10 \\
$A_0$      & -6000 & -6000 \\
\hline
$\mu$          & 200 & 200  \\
$m_A$          & 2000 & 2000 \\
\hline
$m_{\tg}$   & 4852.9 & 5178.7  \\
$m_{\tu_L}$ & 7193.5 & 30256.0 \\
$m_{\tu_R}$ & 7257.6 & 30323.3  \\
$m_{\te_R}$ & 5836.7 & 29961.9  \\
$m_{\tst_1}$& 4008.8 & 1012.0  \\
$m_{\tst_2}$& 5813.7 & 3405.3  \\
$m_{\tb_1}$ & 5809.5 & 3441.5 \\
$m_{\tb_2}$ & 6985.6 & 5336.2  \\
$m_{\ttau_1}$ & 5776.8 & 5639.4  \\
$m_{\ttau_2}$ & 6233.1 & 5772.0  \\
$m_{\tnu_{\tau}}$ & 6237.1 & 5725.2 \\
$m_{\tchi_2^\pm}$ & 1837.1 & 1936.1  \\
$m_{\tchi_1^\pm}$ & 213.8 & 211.3  \\
$m_{\tchi_4^0}$ & 1865.1 & 1938.8  \\ 
$m_{\tchi_3^0}$ & 991.2 & 1018.7  \\ 
$m_{\tchi_2^0}$ & 208.5 & 207.1  \\ 
$m_{\tchi_1^0}$ & 203.3 & 202.2  \\ 
$m_h$       & 123.3 & 124.5 \\ 
\hline
$\Omega_{\tchi_1^0}^{TP}h^2$ & 0.009 & 0.01 \\
$BF(b\to s\gamma)\times 10^4$ & 3.1 & 3.0  \\
$BF(B_s\to \mu^+\mu^-)\times 10^9$ & 3.8 & 3.8 \\
$\sigma^{SI}(\tchi_1^0, p)$ (pb) & $4.3\times 10^{-10}$ & $4.0\times 10^{-10}$ \\
$\sigma^{SD}(\tchi_1^0 p)$ (pb) & $8.0\times 10^{-6}$ & $7.5\times 10^{-6}$  \\
$\langle\sigma v\rangle |_{v\to 0}$  (cm$^3$/sec)  & $1.9\times 10^{-25}$ & $2.0\times 10^{-25}$ \\
$\Delta_{\rm EW}$ & 158 & 25 \\
\hline
\end{tabular}
\caption{Input parameters and masses in~GeV units
  for the NUHM2 and NUHM3 model   SUSY benchmark points
with $m_t=173.2$ GeV using Isajet 7.91.
}
\label{tab:bm}
\end{table}

\section{SUSY parameter space with decoupled first/second generation scalars}
\label{sec:pspace}

\begin{figure}[htb!]
    \centering
    \begin{subfigure}{\textwidth}
        \centering
        \includegraphics[height=0.35\textheight]{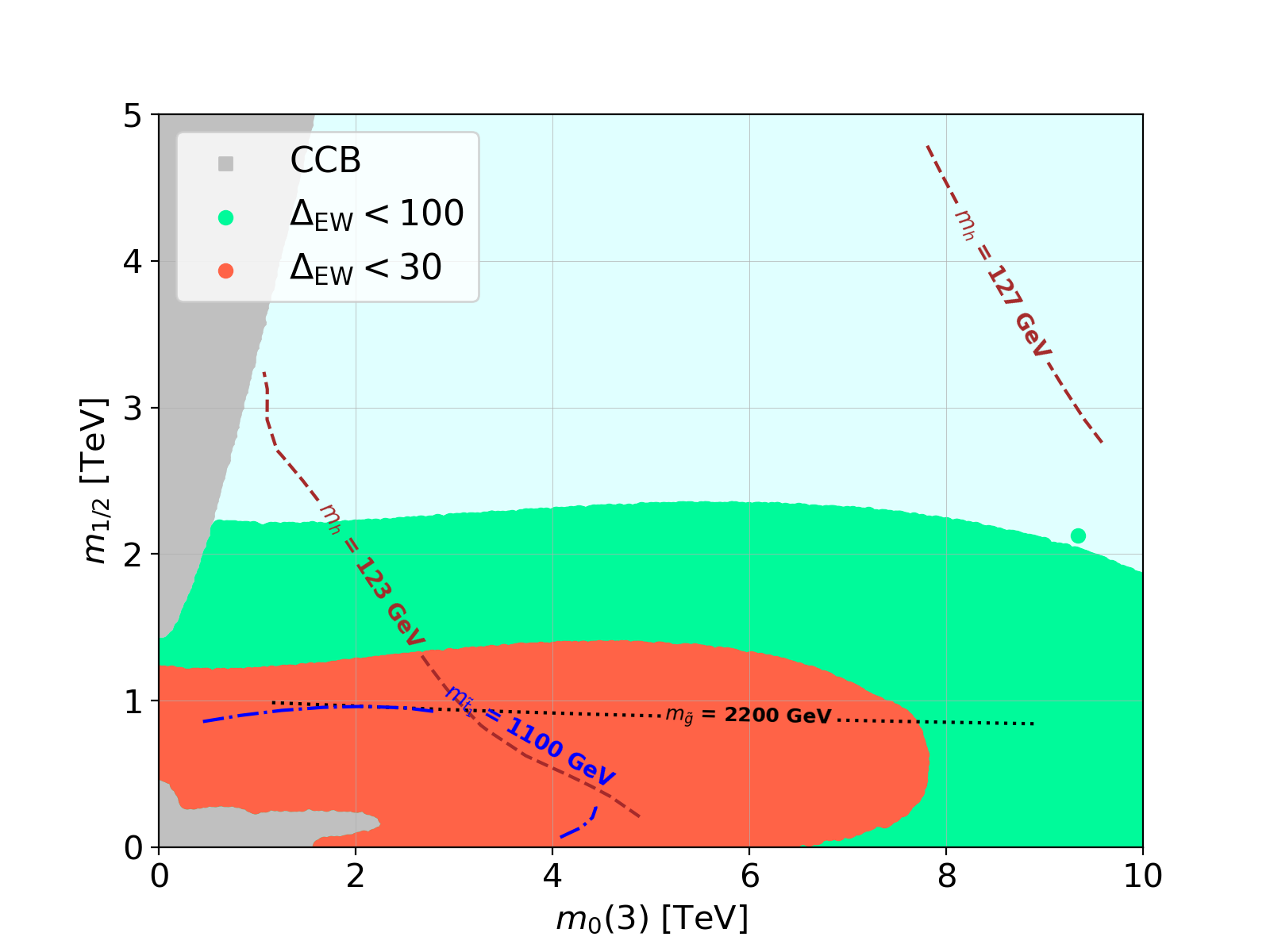}
        \caption{}
    \end{subfigure}
    \begin{subfigure}{\textwidth}
        \centering
        \includegraphics[height=0.35\textheight]{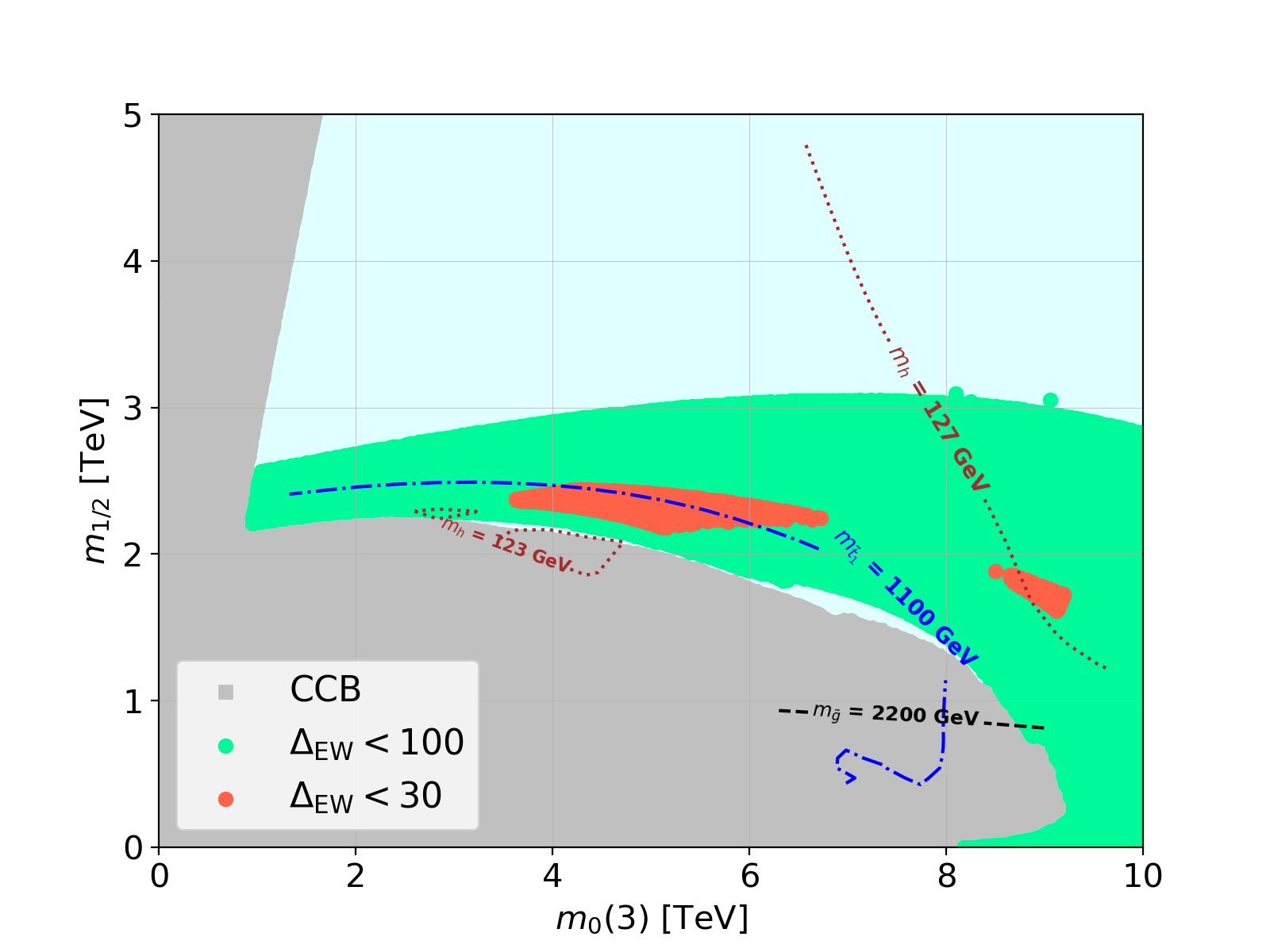}
        \caption{}
    \end{subfigure}
    \caption{The $m_0(3)$ vs. $m_{1/2}$ parameter space of 
          {\it a}) the NUHM2 model with $m_0(1,2)=m_0(3)$ and
          {\it b}) the NUHM3 model with $m_0(1,2)=20$ TeV.
          For both frames, we take $A_0=-1.6 m_0(3)$, $m_A=2$ TeV,
          $\mu =200$ GeV and $\tan\beta =10$. 
                \label{fig:pspace1}}
\end{figure}

In Fig. \ref{fig:pspace1}, we show the $m_0(3)$ vs. $m_{1/2}$ parameter space
plane of {\it a}) the NUHM2 model with $m_0(3)=m_0(1,2)$ and
{\it b}) the NUHM3 model with $m_0(1,2)= 20$ TeV.
For both frames, we take $A_0=-1.6 m_0(3)$, $m_A=2 $ TeV, $\mu =200$ GeV and
$\tan\beta =10$.
In frame {\it a}), the natural region with $\Delta_{EW}\alt 30$ is shown as
red-shaded while green-shaded has $\Delta_{EW}< 100$.
The region between the two orange-dashed contours has $123<m_h<127$ GeV,
so is consistent with LHC $m_h$ measurements.
The region below the purple-dashed contour has
$m_{\tg}< 2.2$ TeV and so is excluded by ATLAS/CMS gluino pair searches
within the context of simplified models\cite{ATLAS:2019vcq,Canepa:2019hph,ATLAS:2024lda}.
The grey-shaded regions contain CCB minima and so are also excluded.
For this NUHM2 case, a large chunk of natural
parameter space is already excluded by LHC searches, but a significant
portion remains natural but with $m_{\tg}> 2.2$ TeV.
Top squarks with mass $m_{\tst_1}<1.1$ TeV lie below the light-blue dotted
contour.
Naively, we would expect this model to suffer from the
SUSY flavor problem if $m_0(3)\ne m_0(1)\ne m_0(2)$, but instead just
average to $m_0(3)$.

As we progress to frame {\it b}) with $m_0(1,2)=20$ TeV, the flavor problem
may be reduced or eliminated.
(Here, the $m_0(1,2)=20$ TeV means just an
average quasi-degeneracy where we might actually expect say $m_0(1)\sim 21$ TeV and $m_0(2)\sim 19$ TeV.)
In this case, the entire lower-left CCB region has
greatly expanded and now we see that LHC $m_{\tg}>2.2$ TeV mass requirements
do not exclude any allowed parameter space with $m_0(3)\alt 9$ TeV.
Still, the bulk of parameter space has $m_h\sim 125$ GeV.
Furthermore, the natural parameter space region has moved up to
$m_{1/2}\sim 1.8 -2.4$ TeV, and shows two disjoint regions.
The left-most larger region is partially excluded by LHC top-squark searches
and has $m_{\tst_1}\alt 1.1$ TeV.
The remainder has top-squark masses just above the LHC bounds.
The smaller right-most natural region has somewhat heavier top-squarks.
From this plot, it is no surprise that LHC hasn't discovered gluinos
with $m_{\tg}\alt 2.2$ TeV since they would likely live in either unnatural
regions of parameter space or else p-space with CCB minima.

\begin{figure}[htb!]
\centering
        {\includegraphics[height=0.4\textheight]{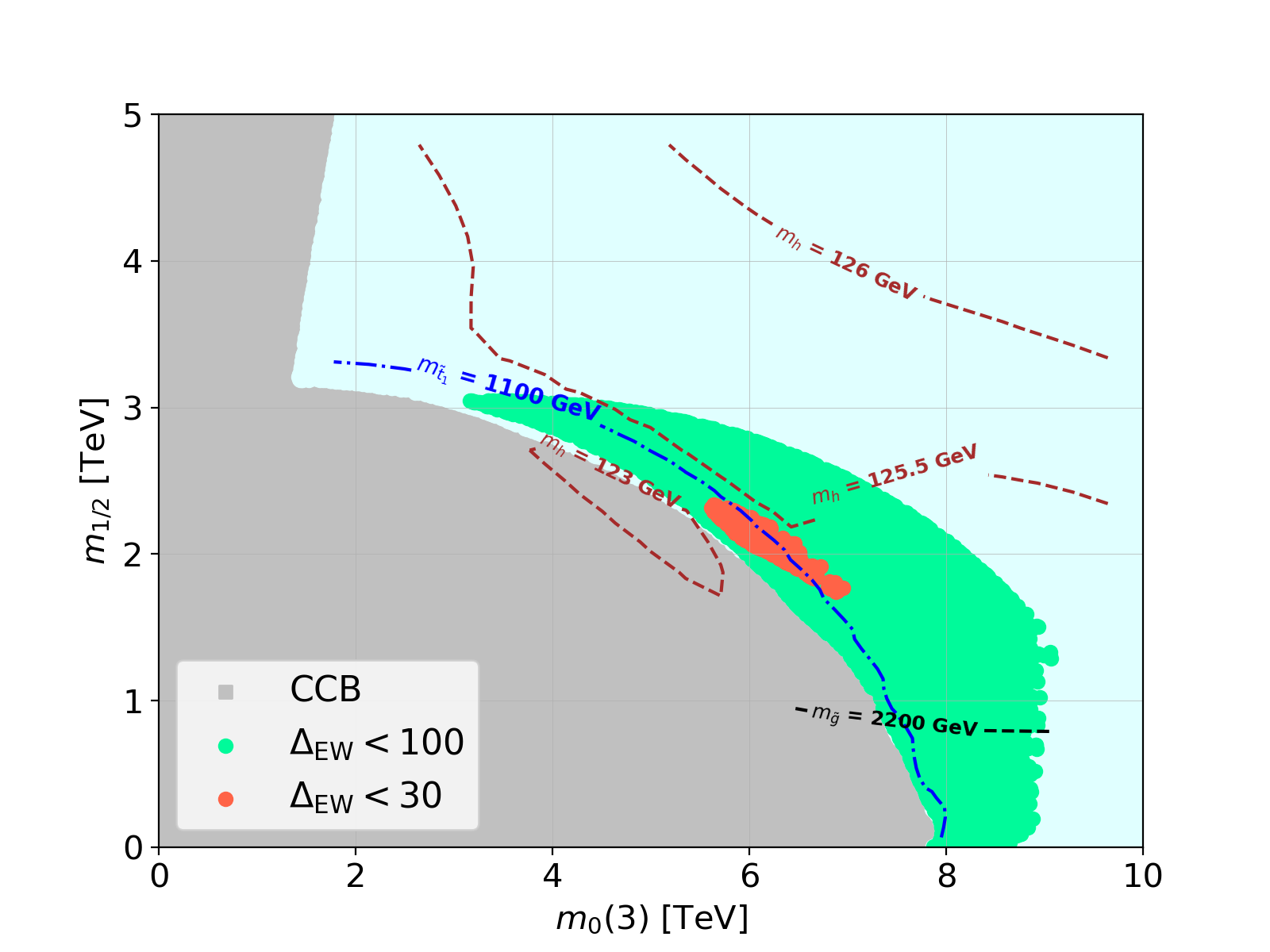}}
        \caption{The $m_0(3)$ vs. $m_{1/2}$ parameters space of 
          the NUHM3 model with $A_0=-m_0(3)$ and $m_0(1,2)=30$ TeV.
          We also take $m_A=2$ TeV, $\mu =200$ GeV and $\tan\beta =10$. 
                \label{fig:pspace2}}
\end{figure}

In Fig. \ref{fig:pspace2}, we show the same NUHM3 parameter space as
Fig. \ref{fig:pspace1} except now pushing the average $m_0(1,2)$ up
to $\sim 30$ TeV. To do this, we must lower $A_0\sim -m_0(3)$.
For this more extreme case, we see the CCB-excluded region has risen as high
as $m_{1/2}\sim 3$ TeV for the left-hand-side of the plot.
Since $m_{\tg}\sim 2.3 m_{1/2}$ here, that corresponds to $m_{\tg}\agt 7$ TeV.
Only one small natural region of p-space remains with $m_0(3)\sim 6$ TeV and
$m_{1/2}\sim 2$ TeV, where $m_{\tg}\sim 4.5$ TeV.
Nonetheless, the value of $m_h\sim 125$ GeV while top-squarks lie not-too-far
from current LHC search limits.
Thus, again if we take the landscape mixed quasi-degeneracy/decoupling
solution to the SUSY flavor problem,
then we should not be surprised that LHC has failed to discover SUSY,
although SUSY could be available within the next $2-10$ years via the 
light higgsino pair production\cite{Baer:2011ec,Han:2014kaa,Baer:2014kya,Baer:2021srt} or top-squark pair production
signatures\cite{Baer:2023uwo}.
This is a consequence of the string landscape draw to large soft terms
which results in living dangerously near the edge of the onset of
CCB minima in the scalar potential.

\begin{figure}[htb!]
\centering
    {\includegraphics[height=0.4\textheight]{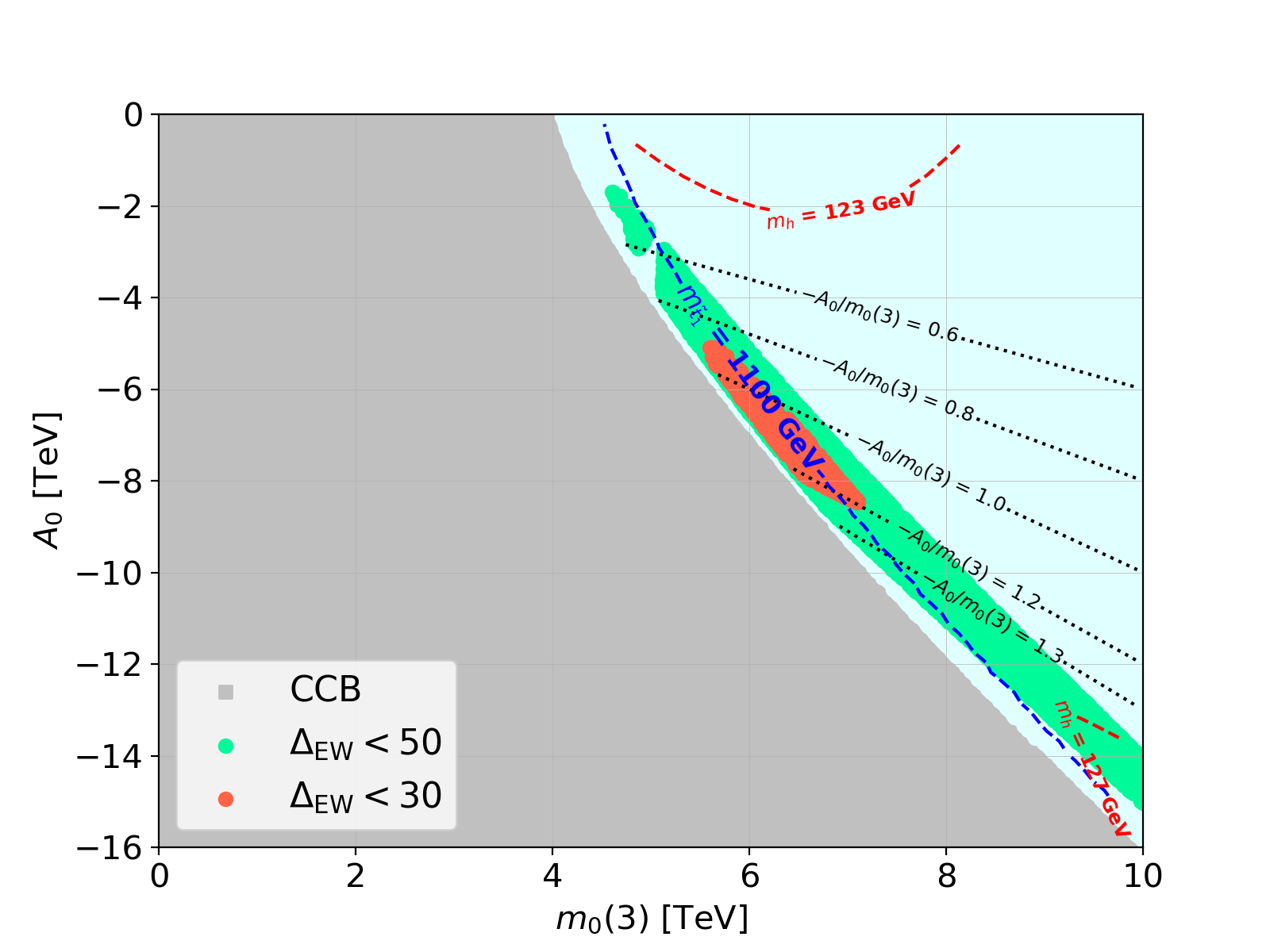}}
        \caption{The $m_0(3)$ vs. $A_0$ parameters space of 
          the NUHM3 model with $m_0(1,2)=30$ TeV.
          We also take $m_{1/2}=2.2$ TeV, $m_A=2$ TeV, $\mu =200$ GeV
          and $\tan\beta =10$. 
                \label{fig:m0A0}}
\end{figure}

In Fig. \ref{fig:m0A0}, we instead show the $m_0(3)$ vs. $A_0$ parameter space
plane in the NUHM3 model with $m_0(1,2)=30$ TeV with $m_{1/2}=2.2$ TeV,
$m_A=2$ TeV and with $\mu =200$ GeV and $\tan\beta =10$.
This plane should again have the landscape mixed decoupling/quasi-degeneracy
solution to the SUSY flavor problem.
In this case, we again see that the red-shaded natural region is rather
small (and hence predictive).
And again: it is living dangerously, rather close to the left gray-shaded
CCB excluded region. The value of $m_h\sim 125$ GeV throughout the bulk of
the allowed parameter space.
For this plot, we also show contours of $A_0/m_0(3)$ since the allowed
natural parameter space is sensitive to
this ratio for large $m_0(1,2)$ in the tens of TeV.
Here, we see that the natural region extends from $-5> A_0>-9$ TeV and
where the ratio $A_0/m_0(3)$ varies between $0.9-1.2$.

\subsection{Top squark mass from the landscape for decoupled first/second gneration sfermions}

\begin{figure}[htb!]
\centering
    {\includegraphics[height=0.4\textheight]{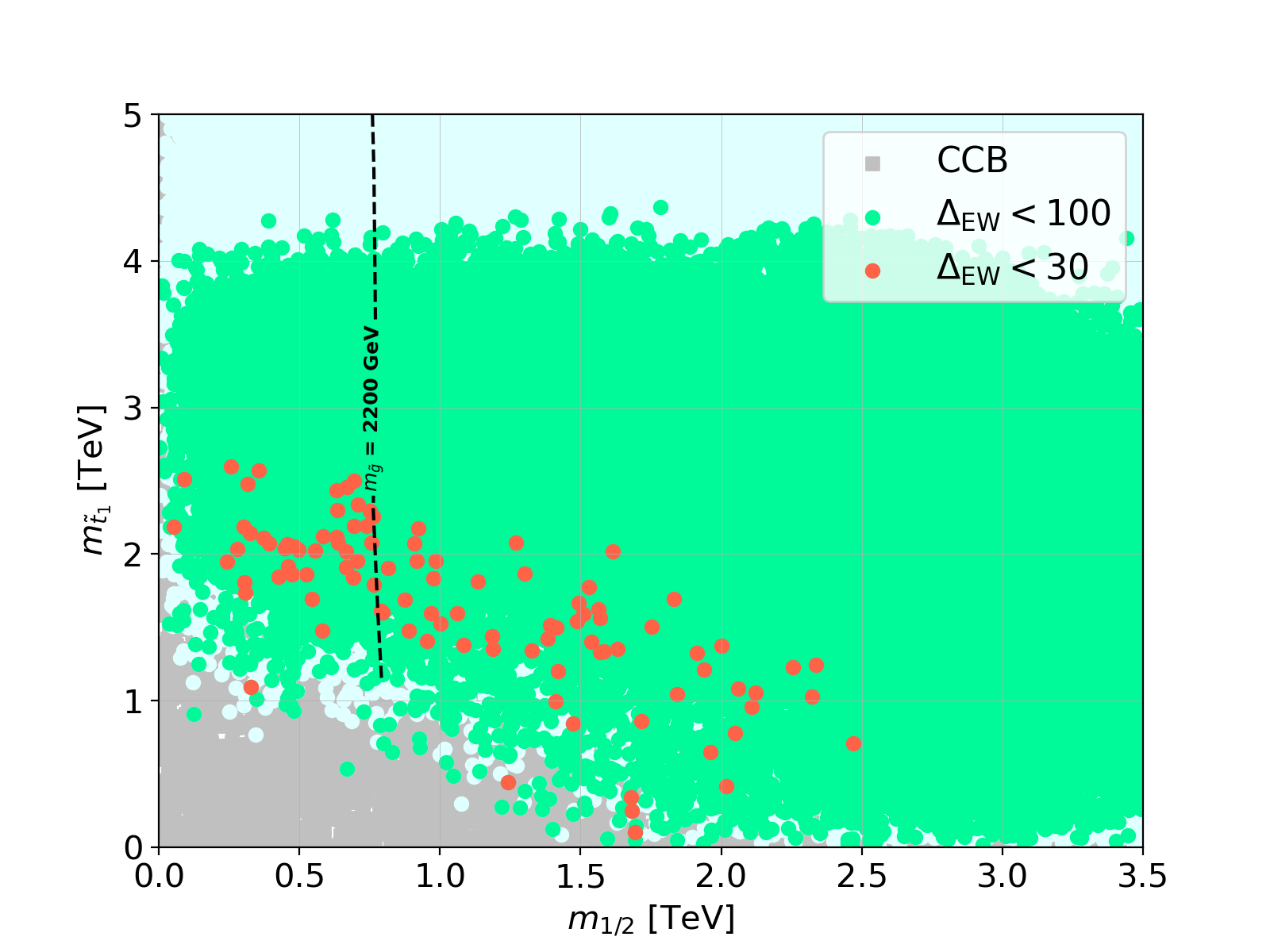}}
    \caption{Distribution of landscape scan points
      in $m_{1/2}$ vs. $m_{\tst_1}$ plane in the NUHM3 model.
            \label{fig:m12vsmt1}}
\end{figure}

To hone our expectations for rather light stops in the $m_{\tst_1}\sim 1-2$ TeV
range,  we return to our landscape scan from Subsec. \ref{ssec:scan}.
From this scan, we plot in Fig. \ref{fig:m12vsmt1} the surviving landscape
scan points in the $m_{1/2}$ vs. $m_{\tst_1}$ mass plane.
Gray dots are excluded as having CCB minima while red dots
lie within the ABDS window with $\Delta_{EW}<30$.
The region to the left of the vertical black contour around $m_{1/2}\sim 0.7$ TeV
has $m_{\tg}\alt 2.2 $ TeV and so is LHC-excluded.
Some of the remaining red dots with $m_{\tst_1}\alt 1.1$ TeV are also excluded
by LHC top-squark searches.
The remaining red dots lie within the $m_{\tst_1}\sim 1-2$ TeV band.
This is to be compared to the recent study of Ref. \cite{Baer:2023uwo}
where prospects for top-squark  detection in natural SUSY models were
examined for HL-LHC. In that work, it was found that 
HL-LHC with $\sim 3$ ab$^{-1}$ could reach to $m_{\tst_1}\sim 1.7-2$ TeV
(lower number is $5\sigma$ while higher number is 95\%CL reach). 

\section{Conclusions}
\label{sec:conclude}

The string landscape offers a mixed decoupling/quasi-degeneracy solution to
the SUSY flavor and CP problems in SUSY models with gravity mediated SUSY
breaking.
The expected spectra is that of radiatively-driven natural SUSY with light
higgsinos $\sim 100-350$ GeV, but with gauginos 
and top squarks in the few TeV and first/second generation sfermions in the
$10-40$ TeV range, while $m_h$ is pushed to $\sim 125$ GeV.
This scenario reconciles naturalness with LHC Higgs and sparticle mass limits
and lack of FCNCs and electric dipole moment (EDM)  deviations.
It also helps explain why no  SUSY signal has so far been seen at LHC:
for decoupled first/second generation sfermions, 
then the bulk of low $m_0(3)$ and $m_{1/2}$ parameter space which would
ordinarily be accessible to LHC13 is actually not available, but is
excluded by the presence of CCB vacua.
The remaining natural parameter space lies adjacent
to the CCB excluded region, providing an example of living dangerously
in the string landscape.
In these favored regions, we expect light higgsinos which may be accessible via 
higgsino pair searches, but also the presence of light top-squarks with
$m_{\tst_1}\sim 1-2$ TeV which should be accessible to HL-LHC.
Gluinos and winos would likely be too heavy to be detected in
this scenario with rather heavy first/second generation sfermions.

{\it Acknowledgements:} 

VB gratefully acknowledges support from the William F. Vilas estate.
HB gratefully acknowledges support from the Avenir Foundation.


\bibliography{highm0}

\begin{thebibliography}{10}
\expandafter\ifx\csname url\endcsname\relax
  \def\url#1{\texttt{#1}}\fi
\expandafter\ifx\csname urlprefix\endcsname\relax\def\urlprefix{URL }\fi
\expandafter\ifx\csname href\endcsname\relax
  \def\href#1#2{#2} \def\path#1{#1}\fi

\bibitem{Murayama:2000dw}
H.~Murayama, {Supersymmetry phenomenology}, in: {ICTP Summer School in Particle
  Physics}, 2000, pp. 296--335.
\newblock \href {http://arxiv.org/abs/hep-ph/0002232}
  {\path{arXiv:hep-ph/0002232}}.

\bibitem{Cremmer:1982en}
E.~Cremmer, S.~Ferrara, L.~Girardello, A.~Van~Proeyen, {Yang-Mills Theories
  with Local Supersymmetry: Lagrangian, Transformation Laws and SuperHiggs
  Effect}, Nucl. Phys. B 212 (1983) 413.
\newblock \href {https://doi.org/10.1016/0550-3213(83)90679-X}
  {\path{doi:10.1016/0550-3213(83)90679-X}}.

\bibitem{Arnowitt:2012gc}
R.~Arnowitt, A.~H. Chamseddine, P.~Nath, {The Development of Supergravity Grand
  Unification: Circa 1982-85}, Int. J. Mod. Phys. A 27 (2012) 1230028,
  [Erratum: Int.J.Mod.Phys.A 27, 1292009 (2012)].
\newblock \href {http://arxiv.org/abs/1206.3175} {\path{arXiv:1206.3175}},
  \href {https://doi.org/10.1142/S0217751X12300281}
  {\path{doi:10.1142/S0217751X12300281}}.

\bibitem{Witten:1981nf}
E.~Witten, {Dynamical Breaking of Supersymmetry}, Nucl. Phys. B 188 (1981) 513.
\newblock \href {https://doi.org/10.1016/0550-3213(81)90006-7}
  {\path{doi:10.1016/0550-3213(81)90006-7}}.

\bibitem{Kaul:1981wp}
R.~K. Kaul, {Gauge Hierarchy in a Supersymmetric Model}, Phys. Lett. B 109
  (1982) 19--24.
\newblock \href {https://doi.org/10.1016/0370-2693(82)90453-1}
  {\path{doi:10.1016/0370-2693(82)90453-1}}.

\bibitem{Baer:2015rja}
H.~Baer, V.~Barger, M.~Savoy, {Upper bounds on sparticle masses from
  naturalness or how to disprove weak scale supersymmetry}, Phys. Rev. D 93~(3)
  (2016) 035016.
\newblock \href {http://arxiv.org/abs/1509.02929} {\path{arXiv:1509.02929}},
  \href {https://doi.org/10.1103/PhysRevD.93.035016}
  {\path{doi:10.1103/PhysRevD.93.035016}}.

\bibitem{Barbieri:1987fn}
R.~Barbieri, G.~F. Giudice, {Upper Bounds on Supersymmetric Particle Masses},
  Nucl. Phys. B 306 (1988) 63--76.
\newblock \href {https://doi.org/10.1016/0550-3213(88)90171-X}
  {\path{doi:10.1016/0550-3213(88)90171-X}}.

\bibitem{Dine:1995ag}
M.~Dine, A.~E. Nelson, Y.~Nir, Y.~Shirman, {New tools for low-energy dynamical
  supersymmetry breaking}, Phys. Rev. D 53 (1996) 2658--2669.
\newblock \href {http://arxiv.org/abs/hep-ph/9507378}
  {\path{arXiv:hep-ph/9507378}}, \href
  {https://doi.org/10.1103/PhysRevD.53.2658}
  {\path{doi:10.1103/PhysRevD.53.2658}}.

\bibitem{Randall:1998uk}
L.~Randall, R.~Sundrum, {Out of this world supersymmetry breaking}, Nucl. Phys.
  B 557 (1999) 79--118.
\newblock \href {http://arxiv.org/abs/hep-th/9810155}
  {\path{arXiv:hep-th/9810155}}, \href
  {https://doi.org/10.1016/S0550-3213(99)00359-4}
  {\path{doi:10.1016/S0550-3213(99)00359-4}}.

\bibitem{Schmaltz:2000gy}
M.~Schmaltz, W.~Skiba, {Minimal gaugino mediation}, Phys. Rev. D 62 (2000)
  095005.
\newblock \href {http://arxiv.org/abs/hep-ph/0001172}
  {\path{arXiv:hep-ph/0001172}}, \href
  {https://doi.org/10.1103/PhysRevD.62.095005}
  {\path{doi:10.1103/PhysRevD.62.095005}}.

\bibitem{Dine:1993np}
M.~Dine, R.~G. Leigh, A.~Kagan, {Flavor symmetries and the problem of squark
  degeneracy}, Phys. Rev. D 48 (1993) 4269--4274.
\newblock \href {http://arxiv.org/abs/hep-ph/9304299}
  {\path{arXiv:hep-ph/9304299}}, \href
  {https://doi.org/10.1103/PhysRevD.48.4269}
  {\path{doi:10.1103/PhysRevD.48.4269}}.

\bibitem{Cohen:1996vb}
A.~G. Cohen, D.~B. Kaplan, A.~E. Nelson, {The More minimal supersymmetric
  standard model}, Phys. Lett. B 388 (1996) 588--598.
\newblock \href {http://arxiv.org/abs/hep-ph/9607394}
  {\path{arXiv:hep-ph/9607394}}, \href
  {https://doi.org/10.1016/S0370-2693(96)01183-5}
  {\path{doi:10.1016/S0370-2693(96)01183-5}}.

\bibitem{Baer:2010ny}
H.~Baer, S.~Kraml, A.~Lessa, S.~Sekmen, X.~Tata, {Effective Supersymmetry at
  the LHC}, JHEP 10 (2010) 018.
\newblock \href {http://arxiv.org/abs/1007.3897} {\path{arXiv:1007.3897}},
  \href {https://doi.org/10.1007/JHEP10(2010)018}
  {\path{doi:10.1007/JHEP10(2010)018}}.

\bibitem{Bagger:1999sy}
J.~A. Bagger, J.~L. Feng, N.~Polonsky, R.-J. Zhang, {Superheavy supersymmetry
  from scalar mass: A parameter fixed points}, Phys. Lett. B 473 (2000)
  264--271.
\newblock \href {http://arxiv.org/abs/hep-ph/9911255}
  {\path{arXiv:hep-ph/9911255}}, \href
  {https://doi.org/10.1016/S0370-2693(99)01501-4}
  {\path{doi:10.1016/S0370-2693(99)01501-4}}.

\bibitem{Baer:2000jj}
H.~Baer, M.~Brhlik, M.~A. Diaz, J.~Ferrandis, P.~Mercadante, P.~Quintana,
  X.~Tata, {Yukawa unified supersymmetric SO(10) model: Cosmology, rare decays
  and collider searches}, Phys. Rev. D 63 (2000) 015007.
\newblock \href {http://arxiv.org/abs/hep-ph/0005027}
  {\path{arXiv:hep-ph/0005027}}, \href
  {https://doi.org/10.1103/PhysRevD.63.015007}
  {\path{doi:10.1103/PhysRevD.63.015007}}.

\bibitem{Arbey:2011ab}
A.~Arbey, M.~Battaglia, A.~Djouadi, F.~Mahmoudi, J.~Quevillon, {Implications of
  a 125 GeV Higgs for supersymmetric models}, Phys. Lett. B 708 (2012)
  162--169.
\newblock \href {http://arxiv.org/abs/1112.3028} {\path{arXiv:1112.3028}},
  \href {https://doi.org/10.1016/j.physletb.2012.01.053}
  {\path{doi:10.1016/j.physletb.2012.01.053}}.

\bibitem{Baer:2014ica}
H.~Baer, V.~Barger, D.~Mickelson, M.~Padeffke-Kirkland, {SUSY models under
  siege: LHC constraints and electroweak fine-tuning}, Phys. Rev. D 89~(11)
  (2014) 115019.
\newblock \href {http://arxiv.org/abs/1404.2277} {\path{arXiv:1404.2277}},
  \href {https://doi.org/10.1103/PhysRevD.89.115019}
  {\path{doi:10.1103/PhysRevD.89.115019}}.

\bibitem{Gabbiani:1996hi}
F.~Gabbiani, E.~Gabrielli, A.~Masiero, L.~Silvestrini, {A Complete analysis of
  FCNC and CP constraints in general SUSY extensions of the standard model},
  Nucl. Phys. B 477 (1996) 321--352.
\newblock \href {http://arxiv.org/abs/hep-ph/9604387}
  {\path{arXiv:hep-ph/9604387}}, \href
  {https://doi.org/10.1016/0550-3213(96)00390-2}
  {\path{doi:10.1016/0550-3213(96)00390-2}}.

\bibitem{Arkani-Hamed:2004ymt}
N.~Arkani-Hamed, S.~Dimopoulos, {Supersymmetric unification without low energy
  supersymmetry and signatures for fine-tuning at the LHC}, JHEP 06 (2005) 073.
\newblock \href {http://arxiv.org/abs/hep-th/0405159}
  {\path{arXiv:hep-th/0405159}}, \href
  {https://doi.org/10.1088/1126-6708/2005/06/073}
  {\path{doi:10.1088/1126-6708/2005/06/073}}.

\bibitem{Arkani-Hamed:2004zhs}
N.~Arkani-Hamed, S.~Dimopoulos, G.~F. Giudice, A.~Romanino, {Aspects of split
  supersymmetry}, Nucl. Phys. B 709 (2005) 3--46.
\newblock \href {http://arxiv.org/abs/hep-ph/0409232}
  {\path{arXiv:hep-ph/0409232}}, \href
  {https://doi.org/10.1016/j.nuclphysb.2004.12.026}
  {\path{doi:10.1016/j.nuclphysb.2004.12.026}}.

\bibitem{Arvanitaki:2012ps}
A.~Arvanitaki, N.~Craig, S.~Dimopoulos, G.~Villadoro, {Mini-Split}, JHEP 02
  (2013) 126.
\newblock \href {http://arxiv.org/abs/1210.0555} {\path{arXiv:1210.0555}},
  \href {https://doi.org/10.1007/JHEP02(2013)126}
  {\path{doi:10.1007/JHEP02(2013)126}}.

\bibitem{Arkani-Hamed:1997opn}
N.~Arkani-Hamed, H.~Murayama, {Can the supersymmetric flavor problem
  decouple?}, Phys. Rev. D 56 (1997) R6733--R6737.
\newblock \href {http://arxiv.org/abs/hep-ph/9703259}
  {\path{arXiv:hep-ph/9703259}}, \href
  {https://doi.org/10.1103/PhysRevD.56.R6733}
  {\path{doi:10.1103/PhysRevD.56.R6733}}.

\bibitem{Bousso:2000xa}
R.~Bousso, J.~Polchinski, {Quantization of four form fluxes and dynamical
  neutralization of the cosmological constant}, JHEP 06 (2000) 006.
\newblock \href {http://arxiv.org/abs/hep-th/0004134}
  {\path{arXiv:hep-th/0004134}}, \href
  {https://doi.org/10.1088/1126-6708/2000/06/006}
  {\path{doi:10.1088/1126-6708/2000/06/006}}.

\bibitem{Douglas:2006es}
M.~R. Douglas, S.~Kachru, {Flux compactification}, Rev. Mod. Phys. 79 (2007)
  733--796.
\newblock \href {http://arxiv.org/abs/hep-th/0610102}
  {\path{arXiv:hep-th/0610102}}, \href
  {https://doi.org/10.1103/RevModPhys.79.733}
  {\path{doi:10.1103/RevModPhys.79.733}}.

\bibitem{Ashok:2003gk}
S.~Ashok, M.~R. Douglas, {Counting flux vacua}, JHEP 01 (2004) 060.
\newblock \href {http://arxiv.org/abs/hep-th/0307049}
  {\path{arXiv:hep-th/0307049}}, \href
  {https://doi.org/10.1088/1126-6708/2004/01/060}
  {\path{doi:10.1088/1126-6708/2004/01/060}}.

\bibitem{Weinberg:1987dv}
S.~Weinberg, {Anthropic Bound on the Cosmological Constant}, Phys. Rev. Lett.
  59 (1987) 2607.
\newblock \href {https://doi.org/10.1103/PhysRevLett.59.2607}
  {\path{doi:10.1103/PhysRevLett.59.2607}}.

\bibitem{Denef:2004ze}
F.~Denef, M.~R. Douglas, {Distributions of flux vacua}, JHEP 05 (2004) 072.
\newblock \href {http://arxiv.org/abs/hep-th/0404116}
  {\path{arXiv:hep-th/0404116}}, \href
  {https://doi.org/10.1088/1126-6708/2004/05/072}
  {\path{doi:10.1088/1126-6708/2004/05/072}}.

\bibitem{Douglas:2004qg}
M.~R. Douglas, {Statistical analysis of the supersymmetry breaking scale} (5
  2004).
\newblock \href {http://arxiv.org/abs/hep-th/0405279}
  {\path{arXiv:hep-th/0405279}}.

\bibitem{Baer:2020vad}
H.~Baer, V.~Barger, S.~Salam, D.~Sengupta, {String landscape guide to soft SUSY
  breaking terms}, Phys. Rev. D 102~(7) (2020) 075012.
\newblock \href {http://arxiv.org/abs/2005.13577} {\path{arXiv:2005.13577}},
  \href {https://doi.org/10.1103/PhysRevD.102.075012}
  {\path{doi:10.1103/PhysRevD.102.075012}}.

\bibitem{Susskind:2004uv}
L.~Susskind, {Supersymmetry breaking in the anthropic landscape}, in: {From
  Fields to Strings: Circumnavigating Theoretical Physics: A Conference in
  Tribute to Ian Kogan}, 2004, pp. 1745--1749.
\newblock \href {http://arxiv.org/abs/hep-th/0405189}
  {\path{arXiv:hep-th/0405189}}, \href
  {https://doi.org/10.1142/9789812775344_0040}
  {\path{doi:10.1142/9789812775344_0040}}.

\bibitem{Arkani-Hamed:2005zuc}
N.~Arkani-Hamed, S.~Dimopoulos, S.~Kachru, {Predictive landscapes and new
  physics at a TeV} (1 2005).
\newblock \href {http://arxiv.org/abs/hep-th/0501082}
  {\path{arXiv:hep-th/0501082}}.

\bibitem{Candelas:1985en}
P.~Candelas, G.~T. Horowitz, A.~Strominger, E.~Witten, {Vacuum configurations
  for superstrings}, Nucl. Phys. B 258 (1985) 46--74.
\newblock \href {https://doi.org/10.1016/0550-3213(85)90602-9}
  {\path{doi:10.1016/0550-3213(85)90602-9}}.

\bibitem{Bae:2019dgg}
K.~J. Bae, H.~Baer, V.~Barger, D.~Sengupta, {Revisiting the SUSY $\mu$ problem
  and its solutions in the LHC era}, Phys. Rev. D 99~(11) (2019) 115027.
\newblock \href {http://arxiv.org/abs/1902.10748} {\path{arXiv:1902.10748}},
  \href {https://doi.org/10.1103/PhysRevD.99.115027}
  {\path{doi:10.1103/PhysRevD.99.115027}}.

\bibitem{Baer:2012cf}
H.~Baer, V.~Barger, P.~Huang, D.~Mickelson, A.~Mustafayev, X.~Tata, {Radiative
  natural supersymmetry: Reconciling electroweak fine-tuning and the Higgs
  boson mass}, Phys. Rev. D 87~(11) (2013) 115028.
\newblock \href {http://arxiv.org/abs/1212.2655} {\path{arXiv:1212.2655}},
  \href {https://doi.org/10.1103/PhysRevD.87.115028}
  {\path{doi:10.1103/PhysRevD.87.115028}}.

\bibitem{Coleman:1973jx}
S.~R. Coleman, E.~J. Weinberg, {Radiative Corrections as the Origin of
  Spontaneous Symmetry Breaking}, Phys. Rev. D 7 (1973) 1888--1910.
\newblock \href {https://doi.org/10.1103/PhysRevD.7.1888}
  {\path{doi:10.1103/PhysRevD.7.1888}}.

\bibitem{Agrawal:1997gf}
V.~Agrawal, S.~M. Barr, J.~F. Donoghue, D.~Seckel, {Viable range of the mass
  scale of the standard model}, Phys. Rev. D 57 (1998) 5480--5492.
\newblock \href {http://arxiv.org/abs/hep-ph/9707380}
  {\path{arXiv:hep-ph/9707380}}, \href
  {https://doi.org/10.1103/PhysRevD.57.5480}
  {\path{doi:10.1103/PhysRevD.57.5480}}.

\bibitem{Baer:2012up}
H.~Baer, V.~Barger, P.~Huang, A.~Mustafayev, X.~Tata, {Radiative natural SUSY
  with a 125 GeV Higgs boson}, Phys. Rev. Lett. 109 (2012) 161802.
\newblock \href {http://arxiv.org/abs/1207.3343} {\path{arXiv:1207.3343}},
  \href {https://doi.org/10.1103/PhysRevLett.109.161802}
  {\path{doi:10.1103/PhysRevLett.109.161802}}.

\bibitem{Baer:2017uvn}
H.~Baer, V.~Barger, H.~Serce, K.~Sinha, {Higgs and superparticle mass
  predictions from the landscape}, JHEP 03 (2018) 002.
\newblock \href {http://arxiv.org/abs/1712.01399} {\path{arXiv:1712.01399}},
  \href {https://doi.org/10.1007/JHEP03(2018)002}
  {\path{doi:10.1007/JHEP03(2018)002}}.

\bibitem{Martin:1993zk}
S.~P. Martin, M.~T. Vaughn, {Two loop renormalization group equations for soft
  supersymmetry breaking couplings}, Phys. Rev. D 50 (1994) 2282, [Erratum:
  Phys.Rev.D 78, 039903 (2008)].
\newblock \href {http://arxiv.org/abs/hep-ph/9311340}
  {\path{arXiv:hep-ph/9311340}}, \href
  {https://doi.org/10.1103/PhysRevD.50.2282}
  {\path{doi:10.1103/PhysRevD.50.2282}}.

\bibitem{Baer:2000xa}
H.~Baer, C.~Balazs, P.~Mercadante, X.~Tata, Y.~Wang, {Viable supersymmetric
  models with an inverted scalar mass hierarchy at the GUT scale}, Phys. Rev. D
  63 (2001) 015011.
\newblock \href {http://arxiv.org/abs/hep-ph/0008061}
  {\path{arXiv:hep-ph/0008061}}, \href
  {https://doi.org/10.1103/PhysRevD.63.015011}
  {\path{doi:10.1103/PhysRevD.63.015011}}.

\bibitem{Baer:2019zfl}
H.~Baer, V.~Barger, D.~Sengupta, {Landscape solution to the SUSY flavor and CP
  problems}, Phys. Rev. Res. 1~(3) (2019) 033179.
\newblock \href {http://arxiv.org/abs/1910.00090} {\path{arXiv:1910.00090}},
  \href {https://doi.org/10.1103/PhysRevResearch.1.033179}
  {\path{doi:10.1103/PhysRevResearch.1.033179}}.

\bibitem{Paige:2003mg}
F.~E. Paige, S.~D. Protopopescu, H.~Baer, X.~Tata, {ISAJET 7.69: A Monte Carlo
  event generator for pp, anti-p p, and e+e- reactions} (12 2003).
\newblock \href {http://arxiv.org/abs/hep-ph/0312045}
  {\path{arXiv:hep-ph/0312045}}.

\bibitem{Baer:2023uwo}
H.~Baer, V.~Barger, J.~Dutta, D.~Sengupta, K.~Zhang, {Top squarks from the
  landscape at high luminosity LHC}, Phys. Rev. D 108~(7) (2023) 075027.
\newblock \href {http://arxiv.org/abs/2307.08067} {\path{arXiv:2307.08067}},
  \href {https://doi.org/10.1103/PhysRevD.108.075027}
  {\path{doi:10.1103/PhysRevD.108.075027}}.

\bibitem{Feng:1999mn}
J.~L. Feng, K.~T. Matchev, T.~Moroi, {Multi - TeV scalars are natural in
  minimal supergravity}, Phys. Rev. Lett. 84 (2000) 2322--2325.
\newblock \href {http://arxiv.org/abs/hep-ph/9908309}
  {\path{arXiv:hep-ph/9908309}}, \href
  {https://doi.org/10.1103/PhysRevLett.84.2322}
  {\path{doi:10.1103/PhysRevLett.84.2322}}.

\bibitem{Kitano:2006gv}
R.~Kitano, Y.~Nomura, {Supersymmetry, naturalness, and signatures at the LHC},
  Phys. Rev. D 73 (2006) 095004.
\newblock \href {http://arxiv.org/abs/hep-ph/0602096}
  {\path{arXiv:hep-ph/0602096}}, \href
  {https://doi.org/10.1103/PhysRevD.73.095004}
  {\path{doi:10.1103/PhysRevD.73.095004}}.

\bibitem{ATLAS:2020dsf}
G.~Aad, et~al., {Search for a scalar partner of the top quark in the
  all-hadronic $t{\bar{t}}$ plus missing transverse momentum final state at
  $\sqrt{s}=13$ TeV with the ATLAS detector}, Eur. Phys. J. C 80~(8) (2020)
  737.
\newblock \href {http://arxiv.org/abs/2004.14060} {\path{arXiv:2004.14060}},
  \href {https://doi.org/10.1140/epjc/s10052-020-8102-8}
  {\path{doi:10.1140/epjc/s10052-020-8102-8}}.

\bibitem{ATLAS:2020xzu}
G.~Aad, et~al., {Search for new phenomena with top quark pairs in final states
  with one lepton, jets, and missing transverse momentum in $pp$ collisions at
  $ \sqrt{s} $ = 13 TeV with the ATLAS detector}, JHEP 04 (2021) 174.
\newblock \href {http://arxiv.org/abs/2012.03799} {\path{arXiv:2012.03799}},
  \href {https://doi.org/10.1007/JHEP04(2021)174}
  {\path{doi:10.1007/JHEP04(2021)174}}.

\bibitem{CMS:2021beq}
A.~M. Sirunyan, et~al., {Search for top squark production in fully-hadronic
  final states in proton-proton collisions at $\sqrt{s} =$ 13 TeV}, Phys. Rev.
  D 104~(5) (2021) 052001.
\newblock \href {http://arxiv.org/abs/2103.01290} {\path{arXiv:2103.01290}},
  \href {https://doi.org/10.1103/PhysRevD.104.052001}
  {\path{doi:10.1103/PhysRevD.104.052001}}.

\bibitem{ATLAS:2019vcq}
{Search for squarks and gluinos in final states with jets and missing
  transverse momentum using 139 fb$^{-1}$ of $\sqrt{s}$ =13 TeV $pp$ collision
  data with the ATLAS detector} (8 2019).

\bibitem{Canepa:2019hph}
A.~Canepa, {Searches for Supersymmetry at the Large Hadron Collider}, Rev.
  Phys. 4 (2019) 100033.
\newblock \href {https://doi.org/10.1016/j.revip.2019.100033}
  {\path{doi:10.1016/j.revip.2019.100033}}.

\bibitem{ATLAS:2024lda}
G.~Aad, et~al., {The quest to discover supersymmetry at the ATLAS experiment}
  (3 2024).
\newblock \href {http://arxiv.org/abs/2403.02455} {\path{arXiv:2403.02455}}.

\bibitem{Baer:2011ec}
H.~Baer, V.~Barger, P.~Huang, {Hidden SUSY at the LHC: the light higgsino-world
  scenario and the role of a lepton collider}, JHEP 11 (2011) 031.
\newblock \href {http://arxiv.org/abs/1107.5581} {\path{arXiv:1107.5581}},
  \href {https://doi.org/10.1007/JHEP11(2011)031}
  {\path{doi:10.1007/JHEP11(2011)031}}.

\bibitem{Han:2014kaa}
Z.~Han, G.~D. Kribs, A.~Martin, A.~Menon, {Hunting quasidegenerate Higgsinos},
  Phys. Rev. D 89~(7) (2014) 075007.
\newblock \href {http://arxiv.org/abs/1401.1235} {\path{arXiv:1401.1235}},
  \href {https://doi.org/10.1103/PhysRevD.89.075007}
  {\path{doi:10.1103/PhysRevD.89.075007}}.

\bibitem{Baer:2014kya}
H.~Baer, A.~Mustafayev, X.~Tata, {Monojet plus soft dilepton signal from light
  higgsino pair production at LHC14}, Phys. Rev. D 90~(11) (2014) 115007.
\newblock \href {http://arxiv.org/abs/1409.7058} {\path{arXiv:1409.7058}},
  \href {https://doi.org/10.1103/PhysRevD.90.115007}
  {\path{doi:10.1103/PhysRevD.90.115007}}.

\bibitem{Baer:2021srt}
H.~Baer, V.~Barger, D.~Sengupta, X.~Tata, {New angular and other cuts to
  improve the Higgsino signal at the LHC}, Phys. Rev. D 105~(9) (2022) 095017.
\newblock \href {http://arxiv.org/abs/2109.14030} {\path{arXiv:2109.14030}},
  \href {https://doi.org/10.1103/PhysRevD.105.095017}
  {\path{doi:10.1103/PhysRevD.105.095017}}.

\end{thebibliography}
\bibliographystyle{elsarticle-num}

\end{document}